%% file: FwdModelingLRG.tex
\documentclass[a4paper,11pt]{article}
\usepackage{jcappub} % for details on the use of the 
\usepackage{newtxtext,newtxmath}
\usepackage[T1]{fontenc}

\DeclareRobustCommand{\VAN}[3]{#2}
\let\VANthebibliography\thebibliography
\def\thebibliography{\DeclareRobustCommand{\VAN}[3]{##3}\VANthebibliography}

\newcommand{\sersic}{S\'ersic}

\usepackage{multirow} % For multirow cells

\let\dingbatcheckmark\checkmark
\let\checkmark\relax
\usepackage{dingbat}
\let\checkmark\dingbatcheckmark

\usepackage{ae,aecompl}

\usepackage{ulem}
\usepackage{orcidlink}
\usepackage{amssymb}
\usepackage{lingmacros}
\usepackage{tree-dvips}
\usepackage{amsmath}
\usepackage{graphicx}
\usepackage{tabularx,booktabs}
\usepackage{dingbat}
\usepackage{diagbox}

\usepackage{subcaption} 
\usepackage{gensymb}
\usepackage{hyperref}
\usepackage{hhline}

\usepackage{bbold}
\graphicspath{ {./fig/} }

\title{Forward modeling fluctuations in the DESI LRGs target sample using image simulations}

\input{authors}

\date{Accepted XXX. Received YYY; in original form ZZZ}

\abstract{We use the forward modeling pipeline, \texttt{Obiwan}, to study the imaging systematics of the Luminous Red Galaxies (LRGs) targeted by the Dark Energy Spectroscopic Instrument (DESI). Imaging systematics refers to the false fluctuation of galaxy densities due to varying observing conditions and astrophysical foregrounds corresponding to the imaging surveys from which \texttt{DESI LRG} target galaxies are selected. We update the \texttt{Obiwan} pipeline, which we previously developed to simulate the optical images used to target DESI data, to further simulate WISE images in the infrared. This addition allows simulating the \texttt{DESI LRGs} sample, which utilizes WISE data in the target selection. Deep DESI imaging data combined with a method to account for biases in their shapes is used to define a truth sample of potential LRG targets. We inject these data evenly throughout the DESI Legacy Imaging Survey footprint at declinations between -30 and 32.375 degrees. We simulate a total of 15 million galaxies to obtain a simulated LRG sample (\texttt{Obiwan LRGs}) that predicts the variations in target density due to imaging properties. We find that the simulations predict the trends with depth observed in the data, including how they depend on the intrinsic brightness of the galaxies. We observe that faint LRGs are the main contributing source of the imaging systematics trend induced by depth. We also find significant trends in the data against Galactic extinction that are not predicted by \texttt{Obiwan}. These trends depend strongly on the particular map of Galactic extinction chosen to test against, implying systematic contamination in the Galactic extinction maps is a likely root cause  (e.g., Cosmic-Infrared Background, dust temperature correction). We additionally observe a morphological change of the \texttt{DESI LRGs} population evidenced by a correlation between OII emission line average intensity and the size of the $z$-band PSF. This effect most likely results from uncertainties in background subtraction. The detailed findings we present should be used to guide any observational systematics mitigation treatment for the clustering of the \texttt{DESI LRGs} sample.
}

\begin{document}

\maketitle

\flushbottom

\label{firstpage}

\section{Introduction}
Modern cosmology is built upon detecting signals from the distant universe. We collect light and gravity waves through various instruments at different frequencies \cite{aghanim2020planck,swetz2011overview,abramovici1992ligo,lacy2020karl,aihara2011eighth}. For galaxy imaging surveys \cite{dark2016dark, laureijs2011euclid, ivezic2019lsst}, in particular, we collect light profiles of galaxies by taking pictures of the night sky. These images help us infer the properties of galaxies, which can be used as fundamental input for model development in Cosmology. However, as light travels through the Milky Way and the Earth's atmosphere to our camera, it is perturbed by different effects. These effects alter the light profile, making the galaxies in the images deviate from their true appearance. We call these effects imaging systematics \cite{rezaie2021primordial, chaussidon2022angular, johnston2021organised, wagoner2021linear}. It is an umbrella term containing any possible effects contributing to galaxy densities' false fluctuation. In practice, we use survey property maps to quantify the strength of different effects on the galaxy densities. These maps include astronomical effects like Galactic dust extinction and stellar density (including stellar streams) and instrument-based effects like image background noise level and seeing.

The correlation between these property maps and the galaxy density indicates how each survey property contaminates our galaxy samples. 

An accurate correction of imaging systematics lays a solid foundation for a stringent and unbiased measurement for all cosmological probes. It is essential for probes that are sensitive to the power spectrum. For example, primordial non-Gaussianity \cite{mueller2022primordial,rezaie2023local,krolewski2023constraining} is sensitive to the large-scale part of the power spectrum. Redshift Space distortion measurements 
\cite{de2021completed} and full shape analysis \cite{philcox2020combining,kobayashi2022full} are sensitive to the overall shape. If not treated properly, imaging systematics would leave an imprint on the galaxy density map that would be degenerate with the primordial non-Gaussianity signal. On the other hand, traditional probes like Baryon Acoustic Oscillations \cite{ross2017clustering} have a characteristic bump when computing their correlation function, thus are less sensitive to imaging systematics; it is possible to obtain a robust measurement even without any treatment of imaging systematics. 

 As ongoing and future galaxy surveys yield increasingly precise data, the margin for error in imaging systematics needs to diminish to achieve greater constraining power. The Dark Energy Spectroscopic Survey (DESI) \cite{DESI2016a.Science,DESI2022.KP1.Instr,DESI2023a.KP1.SV,DESI2023b.KP1.EDR,adame2024desi,Corrector.Miller.2023,FocalPlane.Silber.2023,Snowmass2013.Levi,DESI2016b.Instr} is a Stage-4 \cite{albrecht2006report} cosmology survey that aims at obtaining better cosmological constraints than the former surveys \cite{alam2015eleventh,ahumada202016th}.  DESI is conducted on the Mayall 4-meter telescope at Kitt Peak National Observatory. It is designed to build an unprecedented 3D map of the universe. It will obtain 47 million galaxy and star spectra, and is expected to achieve a precision of less than 0.4\% precision measurement of the BAO distance scale, and 1.05\% precision on the Hubble parameter \cite{vargas2019unraveling}.

Imaging systematics arises from varying observing conditions in imaging surveys. In the case of DESI, the galaxies are selected from the 9th data release (DR9) of the DESI Legacy Imaging Surveys \cite{dey2019overview, LS.dr9.Schegel.2024, SGA.Moustakas.2023}. The survey is comprised of 3 imaging projects on different telescopes: The Beijing-Arizona Sky Survey (BASS) \cite{zou2017project}, the DECam Legacy Survey (DECaLS) \cite{blum2016decam}, and the Mayall z-band Legacy Survey (MzLS) \cite{silva2016mayall}. These images are processed with the DESI image reduction pipeline \textsc{Legacypipe}%
\footnote{\url{https://github.com/legacysurvey/legacypipe}}. \textsc{Legacypipe} transforms these images into a catalog of astronomical objects. 

We developed a forward modeling pipeline, \texttt{Obiwan}%
\footnote{\url{https://github.com/DriftingPig/obiwan\_code/tree/ObiwanLrgPaper}}, to study the effects of imaging systematics. The idea is to add simulated galaxies into real images and extract these simulated galaxies with the image reduction pipeline \textsc{Legacypipe} \cite{lang2010astrometry, hogg2013replacing, lang2020hybrid}. The pixel-to-object transformation is performed in the same way as real DESI galaxies. \texttt{Obiwan} traces the un-cleaned residuals in the image and provides a catalog with the same imaging systematics variation as the tracers we are interested in. \texttt{Obiwan} was first utilized to study the imaging systematics of eBOSS ELGs \cite{kong2020removing, raichoor2021completed}, and it is further developed to fit our purpose here. 

In this work, we use \texttt{Obiwan} to study the imaging systematics trend of \texttt{DESI LRGs}\cite{zhou2023target}: The bright and luminous galaxies selected from DESI Legacy Imaging Surveys. They will be targeted to get spectrum throughout the 5-year DESI main survey. We recover similar systematics trends on simulated LRGs that we call \texttt{Obiwan LRGs}. We also identify spurious trends that are possibly due to imperfect survey property maps or uncertainties in the image calibration stage. The results we present allow a deeper understanding of the sources of artificial variation in the density of \texttt{DESI LRGs}, which we expect will influence how this variation is corrected for in the ultimate clustering analyses of \texttt{DESI LRGs}.

This paper is organized as follows. In Section \ref{sec:data}, we describe the data we used in our analysis. In Section \ref{sec:Pipeline-Description} we describe the pipeline, as well as new updates. Section \ref{sec:target-selection} describes our method to determine the target selection cut. In Section \ref{sec:pre-process}, we discuss the process needed before the simulation runs. In Section \ref{sec:analysis}, we discuss the analysis of output products. In Section \ref{sec:discussion}, we discuss the indication of our results.

\section{Data}
\label{sec:data}
We use three datasets for our work. These datasets originate from the DR9 DESI Legacy Imaging Survey images, and only the image selection criteria are different. The images are compiled from 4 cameras: The Dark Energy Camera (DECam, \cite{decam}) on the Victor M. Blanco 4-meter Telescope at the Cerro Tololo Inter-American Observatory (CTIO); the MOSAIC-3 camera \citep{mosaic3}, formerly mounted on the Nicholas U. Mayall 4-meter Telescope at the Kitt Peak National Observatory; the 90Prime camera \citep{90prime} on the Bok 2.3-meter telescope located on Kitt Peak; and the Wide-field Infrared Survey Explorer (WISE, \cite{wise}) satellite.

The three datasets described below are composed of these images and are processed with the pipeline \textsc{Legacypipe}. The DESI Legacy Imaging Survey DR9 described in Section \ref{sec:sec1-1} uses most of the images, excluding a few defective images. Images were also excluded because the selected region had reached the depth requirement. This set is used to mass produce \texttt{Obiwan LRGs} to study imaging systematics.  The COSMOS Deep catalog in Section \ref{sec:cosmos-deep} utilizes the regions of the sky with many more exposures than a typical region and has much deeper photometry. This set is used to select the truth sample that \texttt{Obiwan} injects into the images. The COSMOS Repeats in Section \ref{sec:cosmos-repeat} uses a large number of observations of the COSMOS field. The images are separated into ten distinct sets, each individually processed. These repeats are used to understand photometric scattering on \texttt{DESI LRGs} to define extended color cut for the injected \texttt{Obiwan LRG-like} sample.

\subsection{DESI Legacy Imaging Survey DR9}
\label{sec:sec1-1}
We use and simulate data from the DR9 DESI Legacy Imaging Survey\footnote{\url{https://www.legacysurvey.org/dr9/description/}} \cite{dey2019overview}, which was used to obtain the sample that DESI uses for target selection and follow-up spectroscopy \cite{chaussidon2023target, hahn2023desi, raichoor2023target, LGRPrelim.Zhou.2020, myers2023target, VIGalaxies.Lan.2023, Spectro.Pipeline.Guy.2023, FBA.Raichoor.2024}. We briefly describe these images here. This imaging survey consists of three optical bands: $g$, $r$, and $z$, and 4 infrared bands from WISE, $W1$, $W2$, $W3$, and $W4$. The optical band is separated into two parts, the South and the North, by a declination (Dec) of $32.375$ in the North Galactic Cap. The South consists of exposures taken from DECam \cite{depoy2008dark}, with a typical seeing of less than 1.3". The North part consists of images from the Beijing-Arizona Sky Survey \citep{bass} using the 90Prime camera for the $g$ and $r$ bands, and the Mosaic z-band Legacy Survey (MzLS) \citep{mzls} for the $z$-band. The BASS survey has a typical seeing of less than 1.7", and MzLS has a less than 1.3" seeing. The pipeline transforms these images into a common grid with a pixel scale of 0.262 arcsec. These pixels are grouped into $3600 \times 3600$ `bricks', and image processing is performed at the individual brick level.

\begin{figure}
\centering
\includegraphics[width=12cm]{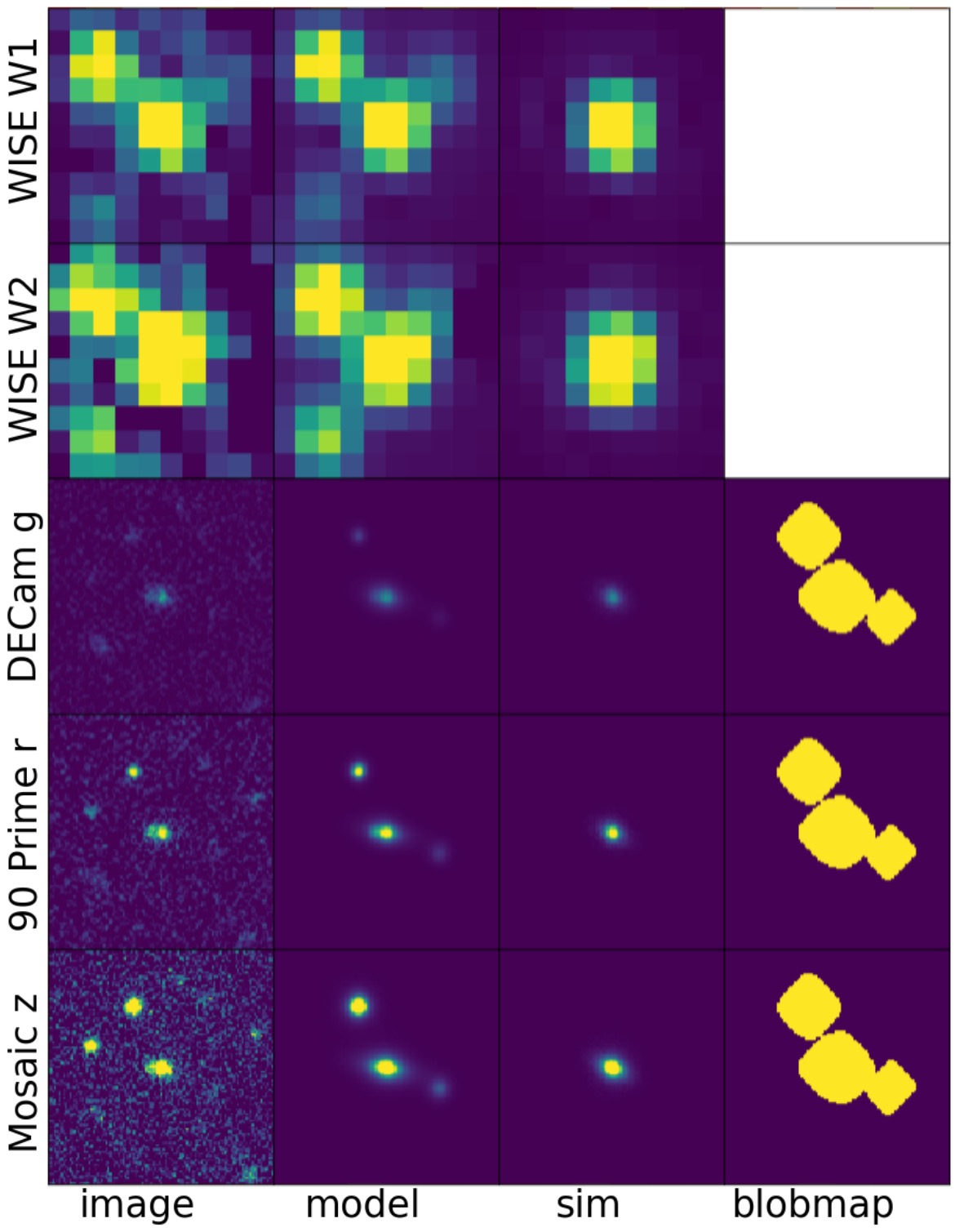}
\caption{Simulated targets in different bands and cameras, processed through the \textsc{Legacypipe} image reduction pipeline. The injected galaxy is located at the center of each image. The figures in the first column are the co-added images. The second column displays the galaxy/star profiles derived from fitting the image. The third column shows the simulated source injected into the images after convolution with the local PSF. The fourth column shows the contiguous area (known as a `blob'), where optical sources within the same blob are fit simultaneously (e.g., to resolve blending). The blobs are not used for WISE processing (see text for details) and, thus, are not shown for their images. The top rows show the WISE $W1$ and $W2$ band images, respectively. The third row shows the DECam image in the $g$- band, and the fourth row shows the $r$- band image from the BASS survey. The third row shows the $z$- band image from the MzLS survey.}
\centering
\label{fig:real-imgs}
\end{figure}

The infrared bands are imaged by the WISE survey \cite{wright2010wide}. WISE is a space telescope whose orbits intersect with its Ecliptic poles ($|\textrm{Dec}|\approx66$) while taking snapshots of the sky during its journey. Because of this orbiting plan, the WISE images have stable orientation in regions far from the Ecliptic poles. Image orientation can be very different during different visits for regions close to the Ecliptic poles. Because of this difference, it is much easier to measure regions far from the Ecliptic poles: The point spread function (PSF) of WISE galaxies is a stable image, and its profile is determined by the orientation of the WISE telescope \cite{lang2014unwise}. Our work focuses only on the Southern part of the imaging survey because we do not have a truth input sample in the northern imaging area, and more details are explained in Section \ref{sec:cosmos-deep}. Though WISE is deeper near the Ecliptic poles, these regions are more challenging to model due to the different image orientations in different visits; however, we avoid these problematic regions since our footprint does not cover that region. The WISE survey has four bandpasses, but only two bands ($W1$ and $W2$) are used in DESI target selection, so we will focus on only $W1$ and $W2$ in what follows. These two bands have a typical point-spread function width of 6.5". The WISE images we process are the unWISE coadded data products \citep{unwise, neo6}, which use a 2.75" pixel scale. Figure \ref{fig:real-imgs} is a representative source with typical PSF size in all bands. It is located in a region observed by all three cameras in the optical bands. The first column shows the real images in optical and infrared bands. The second column shows galaxy models of the real images. The third column shows the galaxy stamp for the simulated galaxy. The fourth column is the ``blob map'': sources within this map are treated as blended sources and will be fit simultaneously. 

\subsection{COSMOS Repeats dataset}
\label{sec:cosmos-repeat}
The COSMOS region has been imaged extensively by many instruments, including DECam.  The COSMOS Repeats dataset we use is a set of catalogs covering $1 \deg^2$ around (RA, Dec) = ($150.1 \deg, 2.2 \deg$). This small region has a large number of exposures taken in all optical bands, which provides the ability to produce many independent subsets of data processed through \textsc{Legacypipe} and containing the same photometric sources. Ten such realizations were produced for the COSMOS repeats data. Gaussian noise was added to each image in each set so that the total \texttt{galdepth}\footnote{It is defined as detection sensitivity of a round exponential galaxy with a radius of 0.45 arcsecond.} in each set was equal to the DESI imaging requirements.  The different image sets are grouped by seeing.
The ten sets are labeled ``set 0'' through ``set 9'', going
from smallest to largest seeing (\texttt{psfsize}). The scatter in photometric measurements of sources common to each set of the COSMOS Repeats data thus provides a fully empirical determination of the expected variance in \textsc{Legacypipe} photometry in different seeing conditions. We simulate sources in the same sets of images with the same settings as COSMOS repeats catalog. We obtained ten sets of corresponding simulated \texttt{Obiwan} COSMOS repeat catalogs that mimic the scattering behavior of sources in real COSMOS repeat catalogs. Achieving consistency with the distributions recovered from the real data in the COSMOS Repeats is our main method of validating the LRG-\texttt{Obiwan} pipeline. The COSMOS Repeats data will also be used to determine our target selection for the extended color boundary that we draw our truth sample from and will be discussed in Section \ref{sec:target-selection}. It will also be used to study PSF error contributions in Section \ref{sec:psfsize}. 

\subsection{COSMOS Deep Catalogs}
\label{sec:cosmos-deep}
The COSMOS Deep Catalogs are image products derived from DECam observations in the COSMOS region, processed with the same \textsc{Legacypipe} procedures as the main DR9 data. The only difference is that the main survey catalog has a maximum depth limit. The depth is closely related to the total number of individual exposures in one processing unit chosen for measurement. Thus, standard Legacy Surveys DR9 release uses far fewer individual images in the COSMOS regions than have been observed by DECam. The Cosmos Deep catalog uses all individual images available and reaches a much higher depth in all optical bands. The higher depth enables the COSMOS deep catalog to reach a better precision in flux and shape measurement, and more detection of faint sources. Details for computational resources used are in Appendix \ref{computational-resources-deep}.  In addition, we perform a cut of \texttt{galdepth\_z} > 24.5 to ensure that the catalog is deep enough in $z$-band. Details about this cut are described in Appendix \ref{appendix:z-band-cut}.

The high depth in optical bands also affects the WISE band: the WISE band images remain the same as in the DR9 release, with a depth much lower than the optical band. The sources in the WISE images are measured in a process called ``forced photometry'': We take the shape and location measured in the optical band, and fit these sources in the WISE images to obtain their flux values in the WISE bands. A higher depth in the optical band has pros and cons for understanding the WISE measurements. It helps improve the accuracy of WISE band fitting by providing more precise information about the galaxies' location and shape. However, since more faint sources are detected in the optical band, the sources in the WISE images are more likely to be blended with other sources. In the presence of blending, the flux on one source could be falsely assigned to a nearby source due to a slight deviation in the shape, PSF, and astrometric position input values. We further test flux accuracy in the WISE band and describe the results in Appendix \ref{appendix:wise-deep}. We find no significant discrepancy in the COSMOS deep catalog for the WISE flux, so we use it as our `truth catalog'.

As the COSMOS Deep data set is imaged with the DECam telescope, the filter properties differ from those used in the BASS/Mzls surveys. Therefore, the COSMOS Deep Catalog is not a representative truth sample for the BASS/MzLS region. Although it is potentially possible to map the truth catalog to the BASS/MzLS region with a filter-transmission function similar to what is described in equation 1-6 of \cite{dey2019overview}, we were unable to perform a robust validation in limited tests. Thus, we focus on the DECaLS data in this study and leave simulating BASS/MzLS for future work. However, given the similarities in the sample selection between the North and South imaging regions, we expect that any conclusions on the properties that drive fluctuations in target density based on our analysis of DECaLS data would hold for BASS/MzLS data.

When sampling from the Deep catalogs, we de-bias the galaxy shapes (this will be discussed in Section \ref{sec:truth-data-generation}). We also apply the LRG-like cut (described in Section \ref{sec:target-selection}), which reduces the sample to only galaxies with a reasonable likelihood of being selected as a DESI target. The final product yields 23,892 sources, with 4,986 that pass the DESI LRG target selection.

\section{The \texttt{Obiwan} Pipeline}
\label{sec:Pipeline-Description}
\begin{figure}
\centering
\includegraphics[width=13cm]{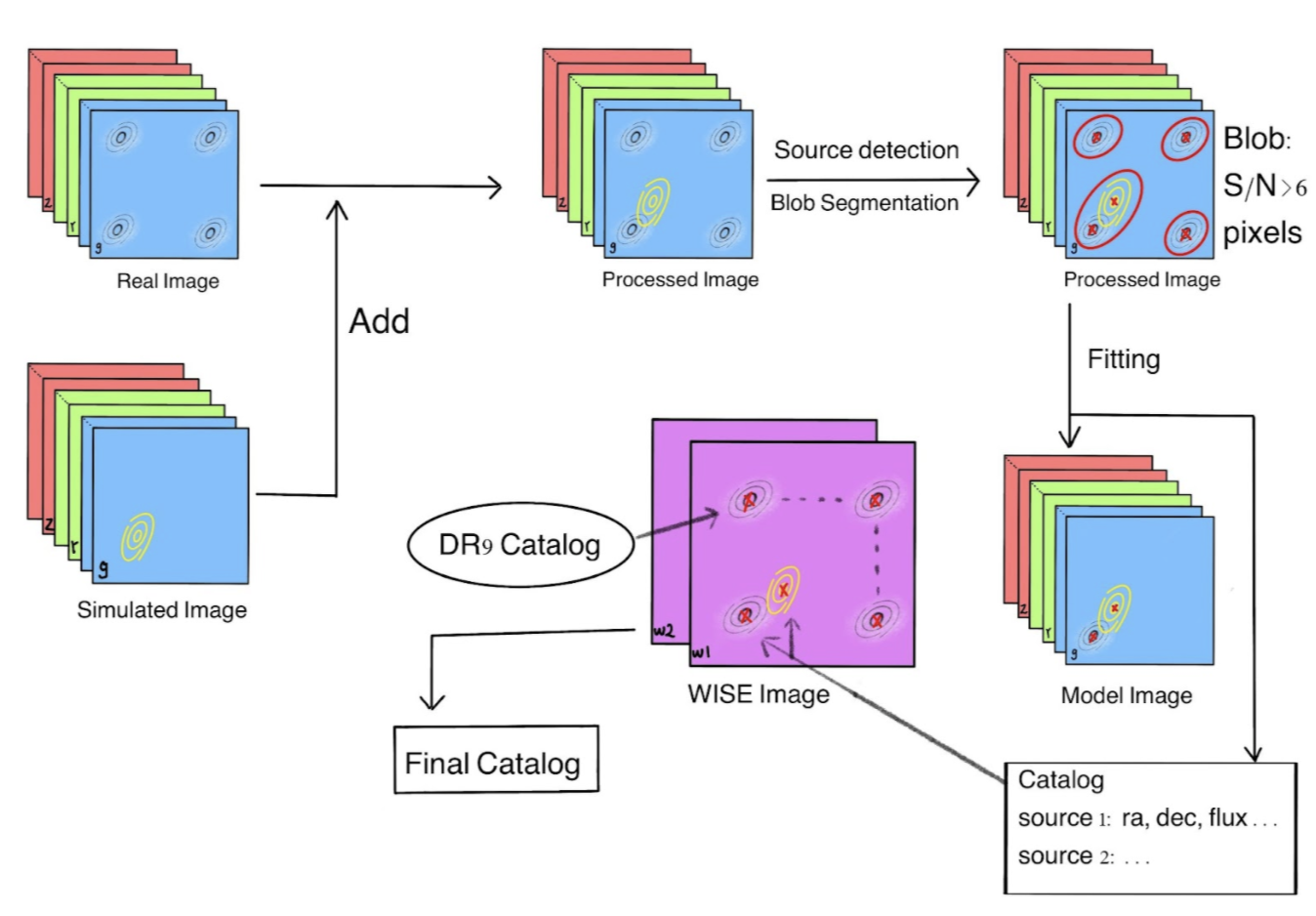}
\caption{Workflow of \texttt{Obiwan}. The blue, green, red, and purple panel represents the $g$, $r$, $z$, $W1$/$W2$ band respectively. The black and white pattern represents the real galaxies, and the yellow pattern represents the simulated galaxies. The workflow shows how the simulated galaxies are added to real images on a per-exposure basis. Only patches with simulated galaxies are fitted. For the fitting of the WISE image, we also need extra information on galaxies from the DR9 data release.}
\label{fig:diagram}
\end{figure}

The synthetic source injection pipeline used in this work is \texttt{Obiwan}. It injects galaxies into the real images, and extracts them using the same procedure applied to real galaxies in the Legacy Surveys DR9 data. This pipeline can be divided into two stages: In the first stage, it injects sources and processes images on optical bands, obtaining the information on flux, shape, etc, for each source. In the second stage, it injects and processes images on the WISE bands. The location and shape of each source, obtained in the first stage, is used in this stage as input. The pipeline fits WISE images to get the WISE flux for each source. Figure \ref{fig:diagram} shows the workflow of this pipeline. 

\subsection{Optical Bands}
\texttt{Obiwan} works by adding simulated sources to real images. The sources are added on a per-exposure basis, at the level of each CCD detector. The RA, Dec location of the injected source is converted into pixel detector units,  the added source is convolved with the PSF at that location, and Poisson noise is applied afterward.  In the \textsc{Legacypipe} processing, these individual exposures are transformed into common brick coordinates. Then, the pipeline proceeds to the source detection stage, where sources are detected and segmented into ``blobs''. A ``blob'' is defined as a set of contiguous pixels with signal-to-noise greater than 6 (when including all images in the optical bands), plus an 8-pixel margin. The blob is defined to be the same region in brick pixel space across all of the optical bands. Each blob will contain at least one source and the photometry is fit independently in each blob. A key aspect that makes \texttt{Obiwan} more efficient is that in the source fitting step, we only fit blobs that overlap the true input RA,Dec of our simulated sources. This helps reduce processing time, and each iteration takes 75\% less time than fitting over the full image. 

\subsection{WISE bands}
Compared to processing a whole image, \texttt{Obiwan} uses much less processing time in the optical band, thanks to only processing a small fraction of sources within the ``blobs'' that have injected galaxies in them. To preserve this structure while computing WISE flux values for \texttt{Obiwan} sources, we need some extra operations on the part of the pipeline that computes WISE flux for galaxies. 

When producing the DR9 catalog, all blobs were processed in optical bands. The WISE band fitting takes all of the source information from optical bands, and uses the information to perform ``forced photometry''. This means that the source location and shape parameters are held fixed and the WISE flux is the only free parameter. ``Forced photometry'' simultaneously fits all sources detected in optical bands over the entire brick of WISE images.

In our \texttt{Obiwan} simulation run, we do not have the optical-band information for all sources. Due to the computationally efficient structure applied in optical bands, we only have the sources within the same blob as simulated sources; sources outside the selected blobs are not processed in previous steps; thus we do not know the details of the sources' shapes and locations in these non-processed blobs. However, these sources are processed and recorded in the DR9 catalogs. These non-processed blobs have conditions identical to those of the sources processed in DR9, so their properties are identical to those recorded in DR9. As a result, we can take these sources' entries in DR9 as input for the WISE flux forced photometry measurement. 

In addition to these sources in blobs that are unaffected by
\texttt{Obiwan} injected galaxies, we must also include bright stars from outside the brick, because bright enough stars will affect the WISE photometry.  We look up and include nearby bright stars in the WISE fitting.  We have validated that the WISE flux derived from this method is identical to a simulation run that processes all blobs in the optical bands.

\section{LRG-like target selection}
\label{sec:target-selection}
We define a new LRG-like cut used to perform target selection on the sample from the COSMOS Deep Catalog. The selected sample with this LRG-like cut served as the truth sample injected into DESI Legacy Imaging Survey DR9 images.

The \texttt{DESI LRGs} sample studied in this work is the LRG SV3 sample described in Table 4 of \cite{zhou2023target}. This selection was used during DESI's Survey Validation phase and is similar to the main LRG target selection, with slightly extended cuts to include more high redshift LRGs. Our work can thus be trivially modified to study LRGs within the main selection cut. 

The sources near selection boundaries can be scattered inside or outside of the LRG target selection boundary. To fully simulate the scattering feature of LRGs, we need to define a \emph{LRG-like color cut}, which spans a larger color selection box than the LRG SV3 selection. At the same time, it also needs to be small enough not to spend too much computing time obtaining the photometry of sources that are unlikely to have measured photometry that matches the LRG SV3 selection. We describe our procedure to generate such a color selection box as follows.

We collect all targets that pass the LRG SV3 selection cuts in the 10 COSMOS Repeats sets. We record the position of all sources, and match these positions to all the other COSMOS Repeats sets.  This new expanded sample is denoted the \emph{Potential-LRGs}. 

Due to photometric scattering, we need to develop a selection scheme that contains LRGs both inside and outside the LRG SV3 color/magnitude selection. We denote this selection scheme \emph{LRG-like} color cut. We use \emph{Potential-LRGs} to quantify the fraction of LRGs inside the \emph{LRG-like} cut: For a given \emph{LRG-like} cut, we define \textbf{scatter rate} as the amount of \emph{Potential-LRGs} outside \emph{LRG-like} cut, divided by the total number of \emph{Potential-LRGs}. We also define \textbf{contamination rate} as the number of sources inside this color cut, divided by the number of \emph{Potential-LRGs} inside the \emph{LRG-like} cut. Our goal for the \emph{LRG-like} cut is to make the selected sample have a low scatter and contamination rates.

\begin{figure}
\centering 
\includegraphics[width=8cm]{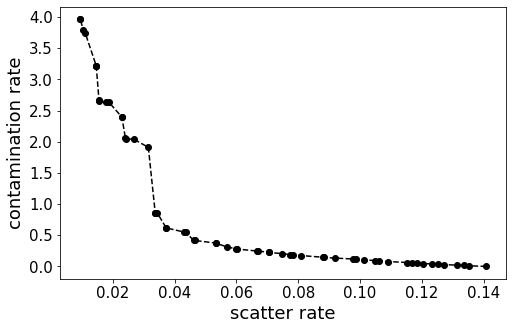}
\caption{The change in scatter and contamination rates as we extend the color selection boundaries. Each point represents a set of color cuts that define the potential LRG-like sample. From the right side to the left side of the plot, we go to a more extended color cut. It would include more potential LRGs, as well as other non-LRGs. We eventually chose the left-most point as our final color cut. This color cut is wide enough to fully mimic the scattering of galaxies outside and inside the LRG SV3 selection boundary while keeping the ratio of input LRGs in \texttt{Obiwan} simulations high enough.}
\label{target_selection}
\end{figure}

The following process describes our method to develop the \emph{LRG-like} cut: We start from the LRG SV3 selection cut, which is composed of 4 equations described in equation \ref{equ:lrg-like-1}--\ref{equ:lrg-like-4} excluding bold numbers. We iteratively extend each selection boundary from equations \ref{equ:lrg-like-1}--\ref{equ:lrg-like-4} by 0.01 magnitudes, and compute the scatter rate for each equation.  We choose the selection cut that corresponds to the lowest scatter rate, and update this selection cut to the new boundary (i.e., increased by 0.01 magnitudes). This would be a new data point in Figure \ref{target_selection}. We perform this process iteratively, widening the selection cuts little by little. Each step is a new realization of a color cut, corresponding to different scatter and contamination rates. Figure \ref{target_selection} shows this variation: Each point represents one potential \emph{LRG-like} color cut. As we extend the selection boundaries, we have a lower scatter rate and a greater contamination rate.

We select the color cut with the lowest scatter rate, as ideally, our simulation would include all sources potentially selected as LRGs. Our final selection corresponds to a scatter rate of 0.009, and a contamination rate of 4.03. These numbers imply that we will include more than 99\% of potential LRG sources and that 20\% of the injected sources will ultimately be selected as LRGs. Follow-up of less than 1\% 

Our final \emph{LRG-like} target selection cut is:

\begin{eqnarray}
\label{equ:lrg-like-1}
z - w1 > 0.8 * (r - z) - 0.6 + \mathbf{0.00}
\end{eqnarray}
\begin{eqnarray}
zfiber < 21.7 + \mathbf{0.32}
\label{equ:fibermagz}
\end{eqnarray}
\begin{eqnarray}
\begin{split}
[(g - r > 1.3 \mathbf{-0.45}) \, \textsc{AND} \\
 (g - r > -1.55 * (r - w1)+3.13 \mathbf{-0.13})] \, \textsc{OR}  \\
(r - w1 > 1.8 \mathbf{-0.37} ) 
\end{split}
\end{eqnarray}
\begin{eqnarray}
\begin{split}
[(r - w1 > (w1 - 17.26) * 1.8 \mathbf{-0.46}) \, \textsc{AND} \\
(r - w1 > (w1 - 16.36) * 1\mathbf{-0.88} )] \, \textsc{OR} \\
(r - w1 > 3.29 \mathbf{-1.65} ) 
\end{split}
\label{equ:lrg-like-4}
\end{eqnarray}

The bold number shows the number of boundaries moved in each color cut. The non-bold part is the original LRG-SV3 selection cut. Eventually, 23892 sources are selected as truth input.

\section{Pre-processing}
\label{sec:pre-process}
\subsection{Truth Data Generation} 
\label{sec:truth-data-generation}
We use the COSMOS Deep dataset to form our input truth set. We apply the LRG-like color cut described in Section \ref{sec:target-selection}, and form a \emph{LRG-like truth set} with clean photometry measured from the Deep data. However, the shape information measured from the COSMOS Deep dataset is not reliable enough for our purposes. This is because the shape measurement in \textsc{Legacypipe} is biased. \textsc{Legacypipe} treats source fitting as a minimization problem, including terms for the chi-squared of the fit, a prior on the ellipticity, and a prior for the model complexity.

\begin{equation}
\label{equ:chi2-model}
\begin{split}
\text{Minimize}\, W = \sum [(\text{image}-\text{model})^2 \times \text{inverse\_variance}]\\
+ \text{ellipticity\_prior}(e_1, e_2)\\
+ \text{parameters\_prior}(N_{\text{parameters}})
\end{split}
\end{equation}

The first term is summed over pixels of a galaxy stamp. We have pixelized data for the observed image (\texttt{image}), the modeled image (\texttt{model}), and a map that indicates the variance for each pixel (\texttt{inverse\_variance}). The second term is a prior that is a function of galaxy ellipticity. The third term increases with the complexity of the galaxy model, and is used to determine the favorable model for a given galaxy. The prior term imposed by ellipticity and the number of parameters is independent of the depth of the image. This means that at higher depth, this prior term becomes less significant since the \texttt{inverse\_variance} term increases and the first term gets bigger than the rest. The \texttt{ellipticity\_prior} term also increases with the intrinsic ellipticity of a galaxy. For galaxies with low \texttt{inverse\_variance}, this helps avoid fitting background noise into the galaxy. However, the measured shape will be biased. This bias also exists in the COSMOS Deep catalog. If we take the shape information as it is, and construct galaxy stamps directly from the COSMOS Deep catalog, the bias in shape would be applied twice: First, the input catalog is biased in shape. Second, our pipeline fitting is the same as \textsc{Legacypipe}, so we would use the same ellipticity prior during the fitting. Our output result made with this method would have a different distribution in shape compared with the real LRGs. We would not be able to fully represent the LRG distribution morphologically. Thus, we develop a procedure to de-bias the shape distribution in the COSMOS Deep truth distribution. The procedure is described in Appendix \ref{appendix:de-bias}. This procedure modifies the input shape of truth catalogs. After this de-biasing procedure, the shape distribution of output \texttt{Obiwan LRGs} is consistent with \texttt{DESI LRGs}.

\subsection{Source injection strategy}
The injected galaxies on the images need to be far from each other while maintaining a large number density. The optimal pattern to achieve this is the hexagonal pattern. This method achieves the maximum number density for a given minimum distance between sources. Thus, we adopt this pattern for source injection.  The distance between the closest sources is 360 pixels, or 94". We deliberately chose this large separation to avoid blending on WISE images. We also eliminate sources that touch pixels containing bright galaxies, bright stars, and clusters; the details are in Appendix \ref{sec:blob-resource}. 

\section{\texttt{Obiwan} with DR9 images}
\label{sec:analysis}

\begin{figure}
    \centering
    \includegraphics[width=9cm]{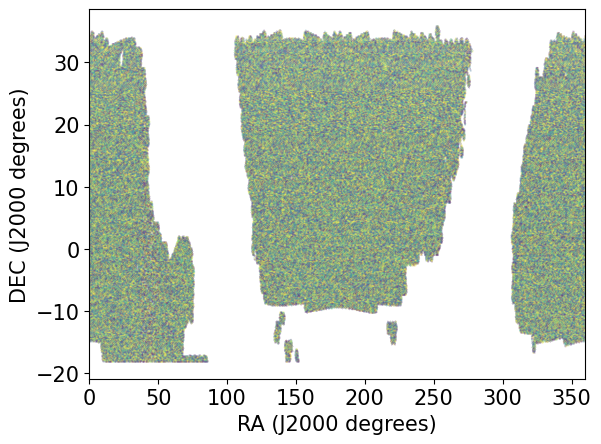}
    \caption{The footprint of \texttt{Obiwan LRGs} we simulated, displayed in Right Ascension and Declination. It is the same footprint as DESI DECaLS NGC and SGC above $\delta=-30$. DESI sources are only targeted above $\delta=-30$, so the region below is not scientifically important. The northern part imaged with BASS/MzLS is not processed, because we do not know the truth input in that region. The patch in the middle is the North Galactic Cap (NGC) region, and the patches on both sides are the South Galactic Cap (SGC) region.}
    \label{fig:den-dr9}
\end{figure}

We inject galaxies in the DECaLS North Galactic Cap (NGC) and South Galactic Cap (SGC) regions, and the footprint is shown in Figure \ref{fig:den-dr9}. If no pixel is masked, 100 galaxies are injected for each brick. In total, 6.6 million galaxies are simulated in the SGC, and 8.5 million galaxies are simulated in the NGC. The density of final \texttt{Obiwan LRGs} is about 30\% that of \texttt{DESI LRGs}. The number density of \texttt{Obiwan LRGs} is limited by computational resources, and we put more details in Appendix \ref{appexdix:resources-obiwan}. We also found that the flux of output galaxies is systematically lower than the injected values. However, this flux bias has a negligible effect on LRG target selection. Appendix \ref{appendix:measurement-bias} provides an in-depth discussion over this issue.

\subsection{Morphology test}
Morphologically, six types of light profiles are included in \textsc{Legacypipe}, and 5 of them are used in our analysis. The unused one is the "DUP" type, a confirmed star by external surveys. Below is a list of the five types of profiles; each type needs a different number of parameters to describe its shape.  
\begin{itemize}
    \item \textbf{PSF}: point source: 0 parameters
    \item \textbf{REX}: round exponential: 1 parameter (radius).  Ellipticity is held constant at zero (circular profile)
    \item \textbf{EXP}: elliptical: 3 parameters (radius and two ellipticity components)
    \item \textbf{DEV}: deVaucouleurs: 3 parameters (radius and two ellipticity components)
    \item \textbf{SER}: \sersic: 4 parameters (\sersic\ index, radius, and two ellipticity components).  The light intensity profile follows $I(R) \varpropto e^{-bR^{1/n}}$, where $R$ is the half-light radius, and $n$ is the \sersic\ index which controls the degree of curvature of the profile. 
\end{itemize}
\begin{figure}
    \centering
    \includegraphics[width=8cm]{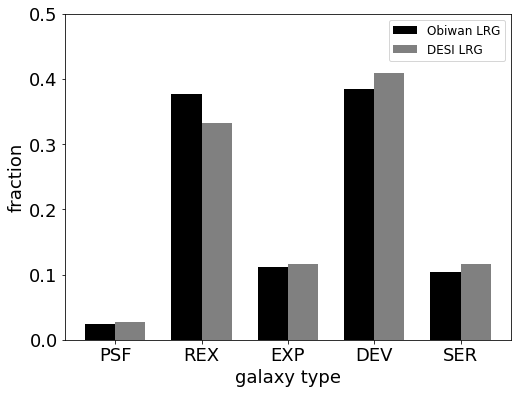}
    \caption{The proportion of different light profiles chosen as the best-fit model in \texttt{DESI LRGs} and \texttt{Obiwan LRGs} in Legacy Surveys DR9. We see that the \texttt{Obiwan LRGs} have similar a proportion in recovered profile types of galaxies as the \texttt{DESI LRGs}. See the text below for the definitions of the labeled profiles.}
    \label{fig:shape-compare}
\end{figure}

\begin{figure}
\centering
\includegraphics[width=12cm]{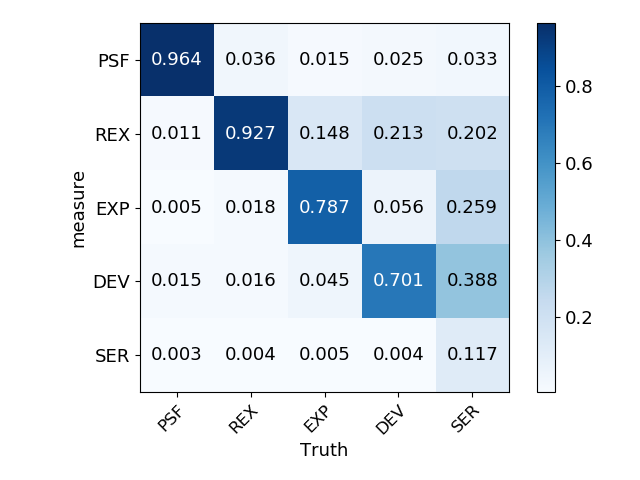}
\caption{A matrix showing the proportion of sources measured given the light profile denoted on the $y$-axis given the input profile used to simulate the source, denoted on the $x$-axis. The fitting pipeline penalizes model complexity and thus will use a more complex model to measure flux only if the image signal to noise is high enough to detect a preference statistically. See the text for details on the specific profile and statistical thresholds.
}
\label{morph}
\end{figure}

Figure \ref{fig:shape-compare} shows the proportion of different types of galaxies in \texttt{DESI LRGs} and \texttt{Obiwan LRGs}. We see good agreement between these two. This results from the debiasing method described in Appendix \ref{appendix:de-bias}, as galaxy types are closely related to galaxy shapes. For example, PSF and REX sources have an ellipticity of 0, so we need the sample to have the right proportion of sources with an ellipticity of 0. With a precise estimate of morphological distribution from the input truth data, we obtain an accurate shape distribution in the output Obiwan LRGs. Figure \ref{morph} shows the scatter between the input and recovered output morphological type of our simulated galaxies. Most galaxies are recovered as their input type, or a simpler type (meaning a model of fewer parameters). As is shown in equation \ref{equ:chi2-model}, the model used depends on several terms. The first term decides how much the image deviates from the model, and it is controlled by the noise level in that location. The third term does not depend on the variance in the image, but only on the number of parameters used in the model. When the image is more noisy, the first term becomes less important, and the pipeline favors a model with fewer parameters. Additionally, there is minimal misclassification of exponential galaxies (EXP) as de Vaucouleurs (DEV) galaxies or vice versa. This could be useful for identifying LRGs with star-forming disks. In Section \ref{sec:psfsize}, we found a correlation between the morphological type of LRGs and $z$-band \texttt{psfsize}. An accurate shape recovery helps us investigate this issue.
%quiescent

\subsection{Imaging Systematics}

We use survey property maps to demonstrate how measurements of galaxy properties are contaminated by numerous effects such as Galactic extinction, denoted as $E(B-V)$; the local density of stars at some flux threshold, denoted as \texttt{stardens}; the detection limit of a point source, denoted as \texttt{psfdepth}\footnote{psfdepth is 5$\sigma$ point source detection limit. \texttt{galdepth} is the 5$\sigma$ detection limit for a fiducial galaxy. They are similar, and here we use \texttt{psfdepth}. The conclusions apply to both.}; and the amount by which the Earth's atmosphere disperses the light we gather with our telescopes, denoted as \texttt{psfsize}. For foreground contaminations like \texttt{stardens} and $E(B-V)$, we use maps \footnote{We specifically used \url{https://data.desi.lbl.gov/public/ets/target/catalogs/dr9/1.1.1/pixweight/main/resolve/dark/pixweight-1-dark.fits}.} produced as described in section 4.5.2 of \cite{myers2023target}. The maps are produced using 
\textsc{HealPix} \cite{Healpix} at a resolution $N_{\rm side}=256$. For effects related to observing conditions, like \texttt{psfsize} and \texttt{psfdepth}, we use direct outputs from the catalog produced by the image reduction pipeline. We see a slight improvement in the matching between \texttt{Obiwan LRGs} and \texttt{DESI LRGs}, particularly in using the catalog level \texttt{psfdepth}, compared with using pixel values in 
\textsc{HealPix}. \texttt{psfdepth} depends on the list of images processed in a certain footprint. The combination of image coadds varies at a resolution smaller than \textsc{HealPix} resolution, so it is also conceptually favorable to test the trends of \texttt{psfdepth} on a catalog level. 

\begin{figure*}
\includegraphics[width=15cm]{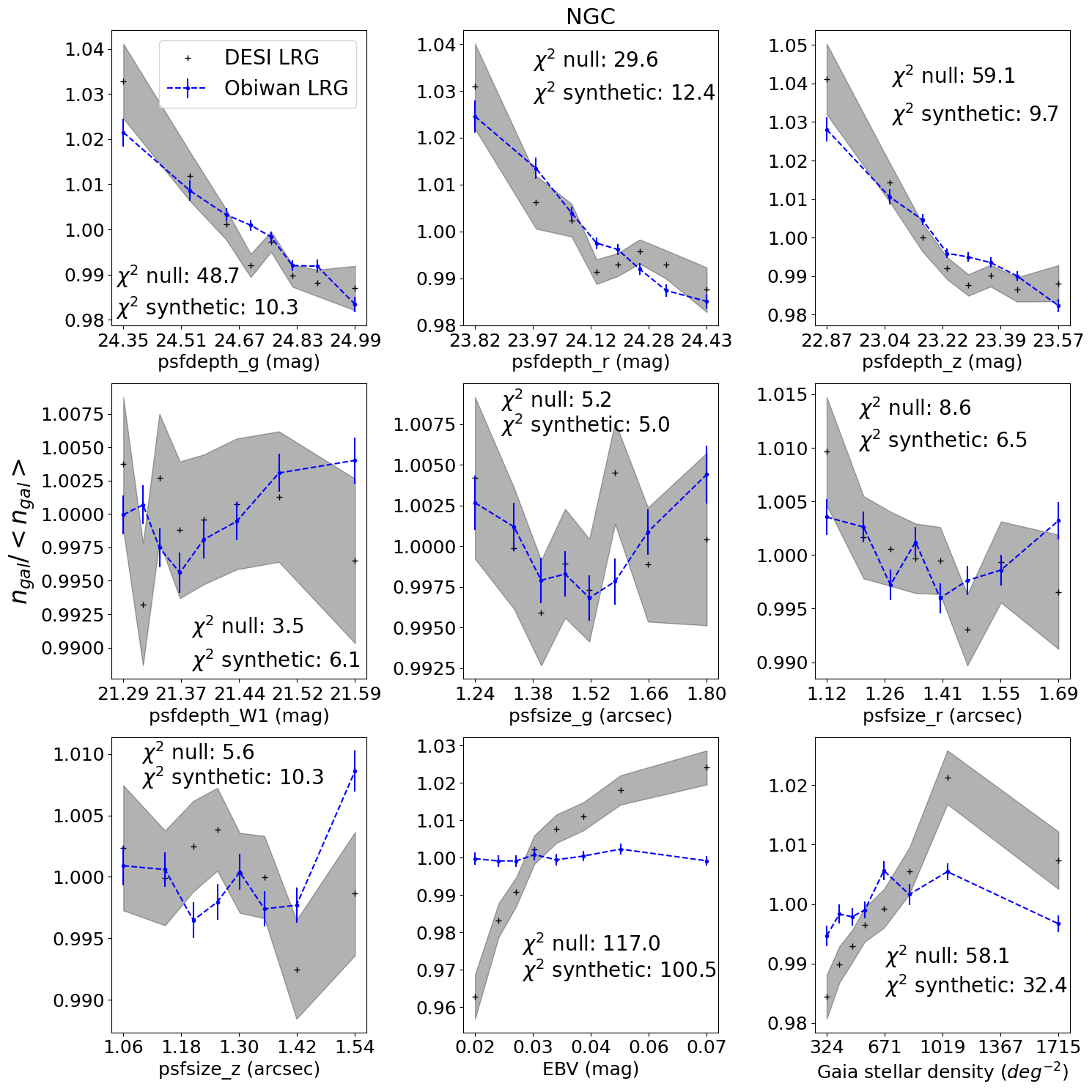}
\caption{Predicted and observed trends between LRG density and imaging properties of the North Galactic Cap. The shaded curves are the results for \texttt{DESI LRGs}, while the blue lines are for synthetic \texttt{Obiwan LRGs}. From top-to-bottom and left-to-right, the first 4 plots are \texttt{psfdepth} trend in $g,r,z,W1$ bands, respectively. The next three plots show the trends against \texttt{psfsize} in the $g,r,z$ bands. Next is the trend against Galactic extinction as measured by \cite{schlegel1998maps},  and finally, the trend against stellar density. The observed trends with \texttt{psfdepth} are predicted. Neither sample has an obvious trend with \texttt{psfsize}. The trends in $E(B-V)$ and stellar density are not predicted by the synthetic Obiwan-LRGs, and we believe this is due to contamination in the $E(B-V)$ map (see Section \ref{sec:$E(B-V)$}).}
\label{fig:systematics-NGC}
\end{figure*}

\begin{figure*}
\includegraphics[width=15cm]{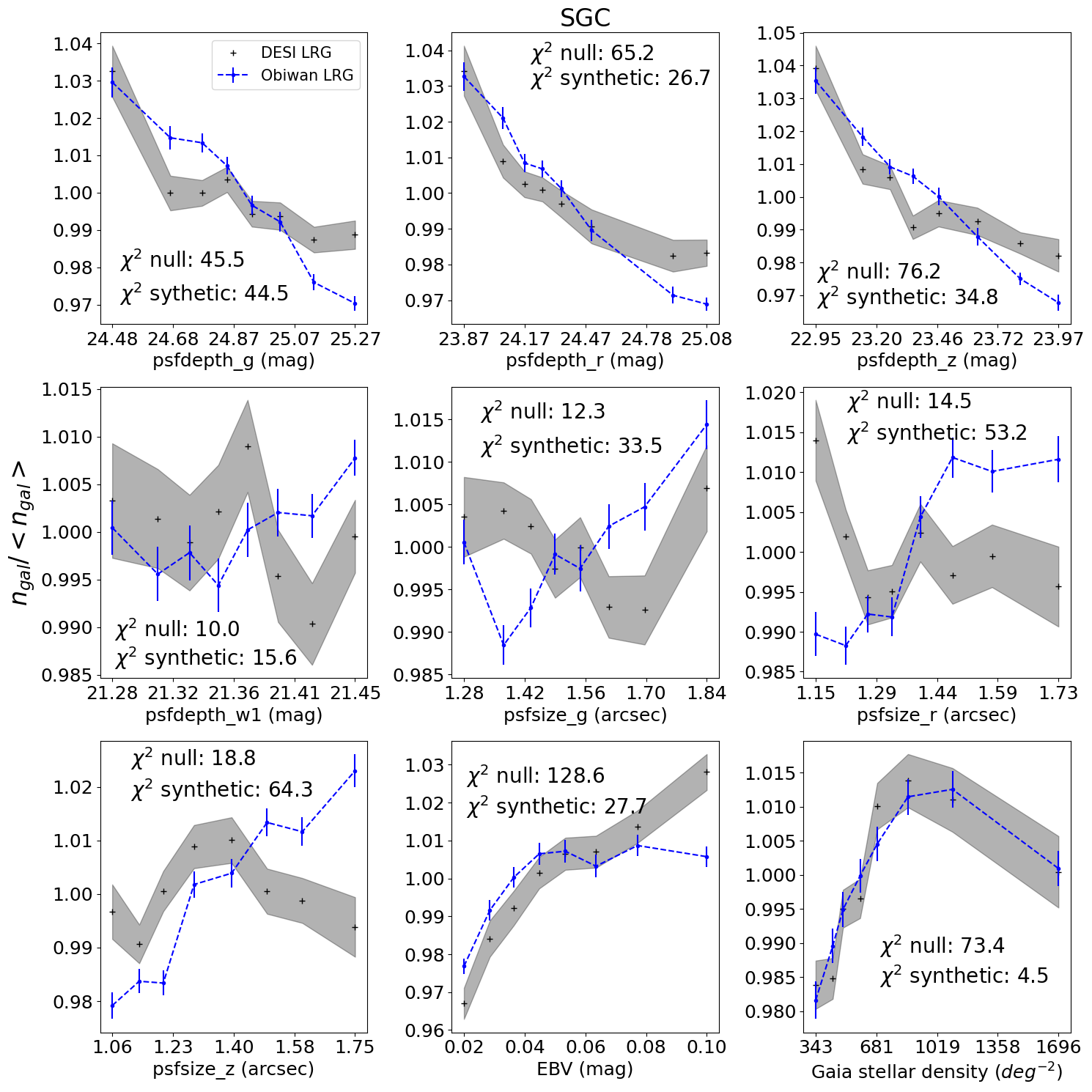}
\caption{The plots are similar to that of Figure \ref{fig:systematics-NGC} except that it is processed on the SGC region. The conclusions are similar to those of NGC. One noticeable difference is that the trends with \texttt{psfdepth} are not as well predicted for the NGC. We believe this is related to how the truth sample is constructed, which induces a mismatch between \texttt{Obiwan LRGs} and \texttt{DESI LRGs}. Details are discussed in Section \ref{sec:correlation-with-galaxy-brigntness}. }
\label{fig:systematics-SGC}
\end{figure*}

We attach the values of \texttt{stardens} ($E(B-V)$ is already in the catalog) from the \textsc{HealPix} map to the catalogs of \texttt{DESI LRGs} and \texttt{Obiwan LRGs}. We use \texttt{Obiwan} input as randoms. Using the \texttt{Obiwan} input as randoms is equivalent to using a general random catalog. The general random catalog is uniformly distributed in the (RA,DEC) sphere. We also apply the same mask to it as \texttt{DESI LRGs}. The random catalog from \texttt{Obiwan input} has the advantage of reducing Poisson noise when tested with \texttt{Obiwan LRGs}. The data counts are binned into eight even percentiles of the given property. Then, the reference random sample is binned in the same way. We divide the data counts by the reference and normalize them by the total number of reference counts to data (so that the overall mean is 1). We obtain results separately for data in the NGC, shown in Figure \ref{fig:systematics-NGC}, and SGC, shown in Figure \ref{fig:systematics-SGC}.

\newcommand{\covlrg}{\mathit{Cov}_{\mathrm{LRG}}}
\newcommand{\covmock}{\mathit{Cov}_{\mathrm{mock}}}
\newcommand{\covobi}{\mathit{Cov}_{\mathrm{Obiwan}}}
\newcommand{\covrand}{\mathit{Cov}_{\mathrm{randoms}}}
\newcommand{\covy}{\mathit{Cov}_{\mathrm{y}}}

To quantify the consistency between the data and the simulated trends via a $\chi^2$ statistic, we first need covariance matrixes for \texttt{DESI LRGs} and \texttt{Obiwan LRGs}. We use:
%how much is Cov_mock larger than Cov_random?

\begin{align}
\covlrg &= \covmock \\
\covobi &= \covmock + \covrand
\end{align}

where the $\covmock$ is obtained from 1000 Flask \cite{xavier2016improving} mocks with the same redshift distribution and density of \texttt{DESI LRGs}. We computed imaging systematics trends for each of the 1000 mocks. For each survey property map $m$, we denote the density variation for mock $i$ as $sys^{m}_{i}$, then the element in row j column k for $Cov^{m}$ (mock or random) is

\begin{equation}
    Cov^{m}_{ik} = \sum_{i=1}^{1000} (sys^{m}_i[j]-sys^{m}_{mean}[j])*(sys^{n}_{i}[k]-sys^{m}_{mean}[k])
\end{equation}

Here $j$, $k$ denotes the bin number of the survey property $m$. We have 8 bins for each survey property map, so j goes from 1 to 8. Similarly, $\covrand$ is obtained from uniformly distributed randoms with the same number density as the \texttt{Obiwan LRGs}, which thus accounts for the noise of their finite statistics. $\covmock$ is much larger than
$\covrand$.

\newcommand{\syslrg}{\mathit{sys}_{\mathrm{LRG}}}
\newcommand{\sysx}{\mathit{sys}_{\mathrm{x}}}

We calculated $\chi^2_{\rm Obiwan}$ values comparing the trends observed in the real LRGs compared to the Obiwan LRGs
\begin{equation}
\begin{split}
\chi^2_{\rm Obiwan} = 
(\syslrg - sys_{\rm Obiwan}+c\cdot \mathbb{1}) \cdot \covobi \cdot  \\
(\syslrg - sys_{\rm Obiwan}+c\cdot \mathbb{1})^T
\end{split}
\end{equation}
and also compared to the null expectation of 1,
\begin{equation}
\chi^2_{\rm null} = (\syslrg-\mathbb{1}+c\cdot \mathbb{1}) \cdot \covlrg \cdot  (\syslrg -\mathbb{1}+c\cdot \mathbb{1})^T
\end{equation}
Here the constant \textbf{c} shifts systematic trend difference up or down to minimize the $\chi^2$ value, $sys_{lrg}$ and $sys_{Obiwan}$ are the systematic trend in \texttt{DESI LRGs} and \texttt{Obiwan LRGs}. The constant $c$ can be theoretically calculated as:
\begin{equation}
c = \frac{-(\syslrg - \sysx) \cdot \covy \cdot \mathbb{1}}{\mathbb{1} \cdot \covy \cdot \mathbb{1}}
\end{equation}

\noindent Here $x$ is \texttt{Obiwan} or \emph{null}, and $y$ is \texttt{Obiwan} or \emph{LRG}.

With a degree of freedom of 8, a trend that fully resembles \texttt{DESI LRGs} trend should have a $\chi^2<13.4$ at 90\% confidence level, or $\chi^2<15.5$ at 95\% confidence level. In the NGC, the plot shows that there is no clear systematic trend in \texttt{psfsize g}, \texttt{psfsize r}, \texttt{psfsize z}, and \texttt{psfdepth W1}, either predicted or observed. We see that the Obiwan LRGs successfully predict the trends observed in \texttt{psfdepth g}, \texttt{psfepth r}, and \texttt{psfdepth z}. The images' depth indicates the galaxies' noise level, which suggests that the systematic trends are from photometric scattering. However, we do see trends in extinction and stellar density that are not predicted by \texttt{Obiwan}. We investigate these further in Section \ref{sec:$E(B-V)$}.

In the SGC, we see qualitative agreement in the slopes of the trends, but the $\chi^2$ results suggest a poor match between the \texttt{DESI LRGs} trends and those that \texttt{Obiwan} predicts. The mismatch for \texttt{psfdepth} is pronounced in high-depth regions. We explain the reason for this mismatch in Section \ref{sec:correlation-with-galaxy-brigntness}. In short, this mismatch does not result from failure in \texttt{Obiwan} simulation. It arises from not having a perfect truth input. The mismatch can be corrected if we match the color distribution of \texttt{DESI LRGs} with \texttt{Obiwan LRGs}.

\section{Testing the \texttt{Obiwan} procedure}
\label{sec:discussion}
There are several trends between the observed LRG density and imaging properties that we could not simulate in the fiducial \texttt{Obiwan} settings. In this section, we discuss components not simulated by \texttt{Obiwan}, which causes a mismatch in imaging systematics trends that we see in Figure \ref{fig:systematics-NGC} and \ref{fig:systematics-SGC}. Section \ref{sec:correlation-with-galaxy-brigntness} discusses the dependency of imaging systematics on intrinsic galaxy brightness, and how imperfections in the "truth catalog" cause mismatch by not having the perfect magnitude distribution. Section \ref{sec:$E(B-V)$} discusses uncertainties in the $E(B-V)$ maps and its correlation with the number density of \texttt{DESI LRGs}. Section \ref{sec:psfsize} discusses signs of un-simulated PSF error and that it could potentially change the characteristics of the LRG population near the color selection boundaries. Section \ref{sec:other_effects} discusses other non-simulated secondary effects that could bias our results, but we do not see a clear sign of their impact. Section \ref{sec:imgsys-discuss} discusses potential issues with using survey property maps for imaging systematics mitigation, the most widely used method for current galaxy surveys. We identify two factors not commonly considered with this method and how forward modeling tools like \texttt{Obiwan} can help. Section \ref{sec:cosmology-analysis} talks about how the systematics discussed in this work impact cosmological measurements by potentially introducing a varying galaxy bias that is not properly mitigated.  

\subsection{Correlation With Galaxy Brightness}
\label{sec:correlation-with-galaxy-brigntness}
\begin{figure}
\centering
\includegraphics[width=\linewidth]{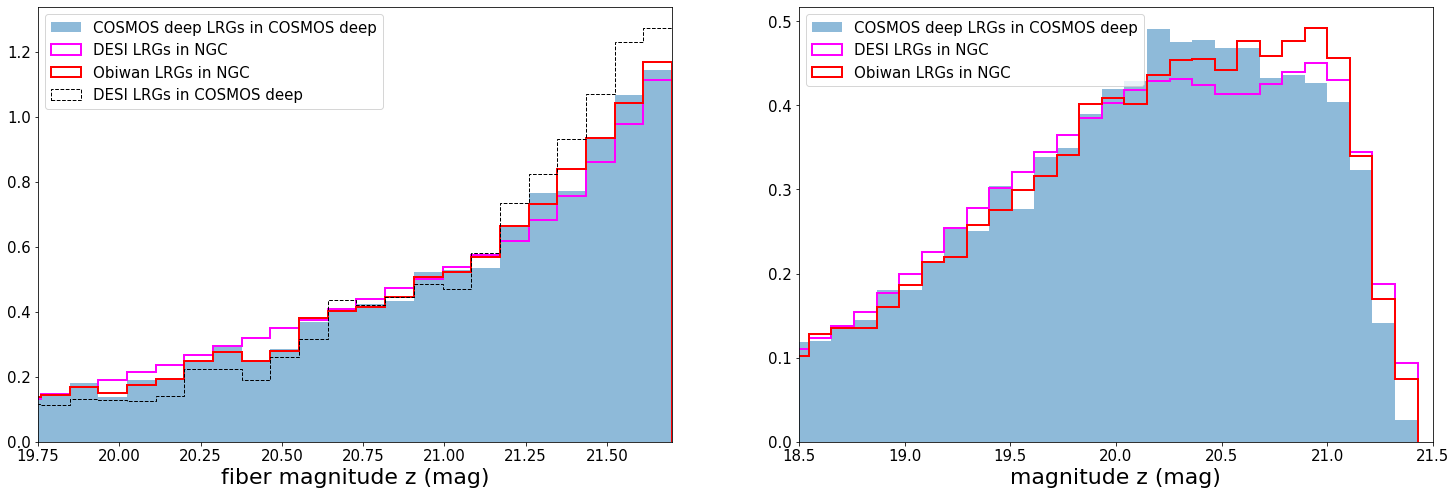}
\caption{The $z$-band fiber magnitude and total magnitude distributions of LRGs in different catalogs and regions. The blue histograms show LRGs in the COSMOS Deep catalog.  The black histogram shows \texttt{DESI LRGs} cut to the COSMOS Deep footprint. The magenta histograms show \texttt{DESI LRGs} in the whole North Galactic Cap (NGC) region.  The red histograms show \texttt{Obiwan LRGs} in the NGC region. The left plot shows a histogram in $z$-band fiber magnitude. The right plot shows the total $z$-band magnitude. The left plot suggests that the COSMOS Deep region is a small region, and \texttt{DESI LRGs} in this same region (Black) have a very different distribution compared to \texttt{DESI LRGs} in a much larger footprint (Magenta). However, LRGs selected from the COSMOS Deep catalog (Blue) show a better resemblance to the histogram of LRGs in a larger footprint (Magenta). The COSMOS Deep LRGs (Blue) have an under-density feature at around $z$-band fiber magnitude $20.5$. This feature is transferred to \texttt{Obiwan LRGs} (Red) with \texttt{Obiwan} simulation so we see the same under-density in \texttt{Obiwan LRGs} in the same $z$-band fiber magnitude range.}
\label{fig:z-deep-dr9}
\end{figure}

The trends with \texttt{psfdepth} still have a small discrepancy between what is observed and predicted, especially at the greatest depths. This originates from the small mismatch in the magnitude distribution between \texttt{DESI LRGs} and \texttt{Obiwan LRGs}, shown in Figure \ref{fig:z-deep-dr9}. The panels show that \texttt{Obiwan LRGs} are generally fainter than \texttt{DESI LRGs}, especially in the $z$-band. We investigate possible explanations for this mismatch.

One possible reason is that the COSMOS Deep data is not strictly noiseless. Although COSMOS Deep is overall a clean sample in optical bands, the W1 flux in the WISE band does not have deep photometry, as is shown in Figure \ref{fig:depth_compare}. The luminosity cut shown in equation \ref{equ:lrg-like-4} is an important source of photometric scattering, and it involves utilizing the WISE W1 band. We do not find evidence of any strong discrepancies induced by imperfect WISE band input, but we can not rule out the possibility that it could cause a mismatch in color distribution in other bands.

A second reason is the sample variance from the truth catalog. The truth catalog is from a small footprint, so the LRG-like sample selected in this region is slightly different from the LRG-like sample in the full footprint. This sample from the truth catalog serves as seeds that generate \texttt{Obiwan} inputs. We inject these \texttt{Obiwan inputs} to images and further produce \texttt{Obiwan LRGs}. Therefore, the features exclusively of the LRGs from the truth catalog can also be seen in \texttt{Obiwan LRGs}. We demonstrate this effect in Figure \ref{fig:z-deep-dr9}. In the left plot, we see LRGs selected from the COSMOS deep catalog have a dip at $z$-band fiber magnitude around 20.5. As we use COSMOS deep catalog as "truth input", this feature is also seen in \texttt{Obiwan LRGs}. Due to this sample variance, the COSMOS deep region LRGs also have a higher density LRGs at the fainter part of $z$-band fiber magnitude (>21.4) than the overall magnitude distribution of \texttt{DESI LRGs}. 

Thus, our discrepancy comes from not having a large enough footprint of the COSMOS Deep catalog. The mismatch does not appear only on $z$-band, but compared with other bands, $z$-band is easier to see the difference by eye. Thus, we choose $z$-band data for illustration. The result here suggests that we should use a larger footprint to generate the ``truth catalog'' in future work. If everything worked correctly, we would see a better match between \texttt{DESI LRGs} in COSMOS deep (or any other deep data used), and the \texttt{DESI LRGs} in the NGC (or any other larger footprint). We should also see a much smaller difference between the $z$-band magnitude distribution of \texttt{Obiwan LRGs} and \texttt{DESI LRGs}.

\begin{figure*}
    \centering
    \includegraphics[width=.85\linewidth]{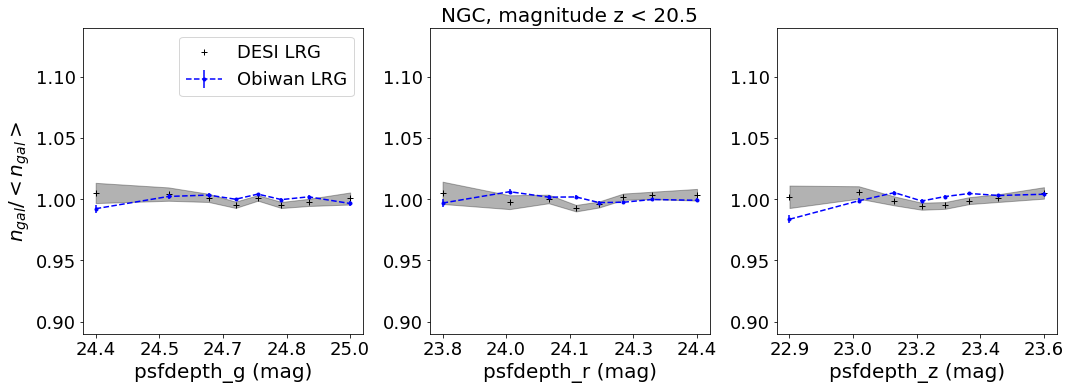}
    \centering
  \includegraphics[width=.85\linewidth]{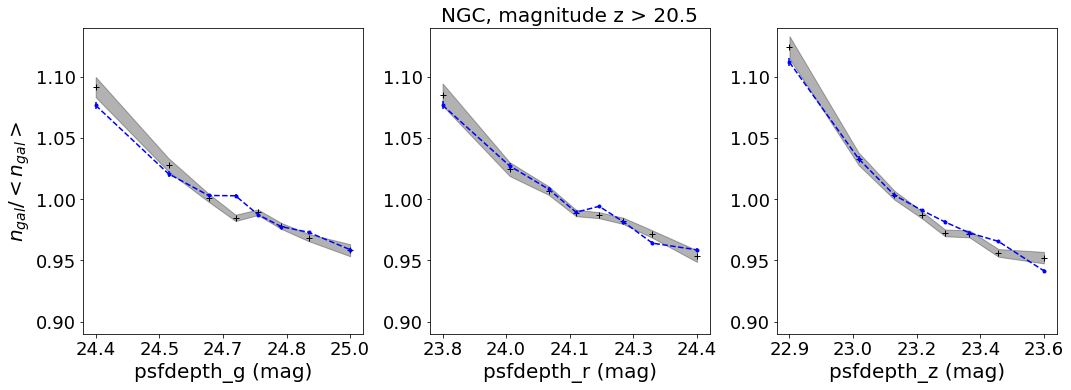}
  %\caption{A subfigure}
  \includegraphics[width=.85\linewidth]{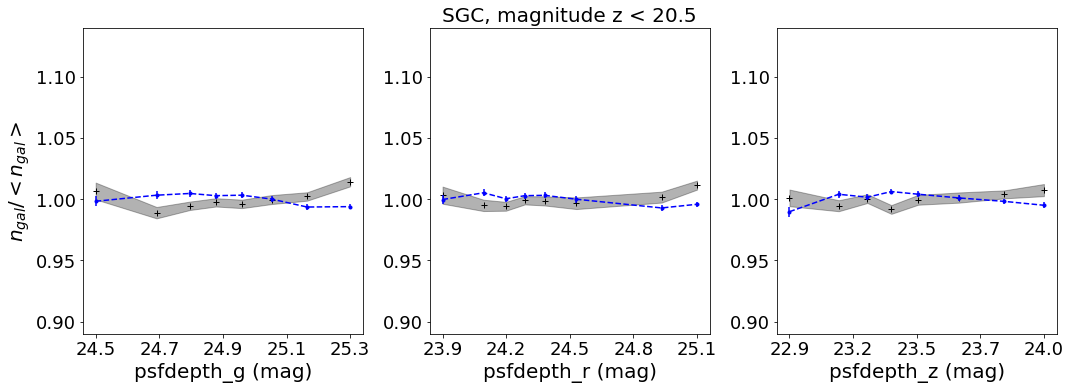}
  \includegraphics[width=.85\linewidth]{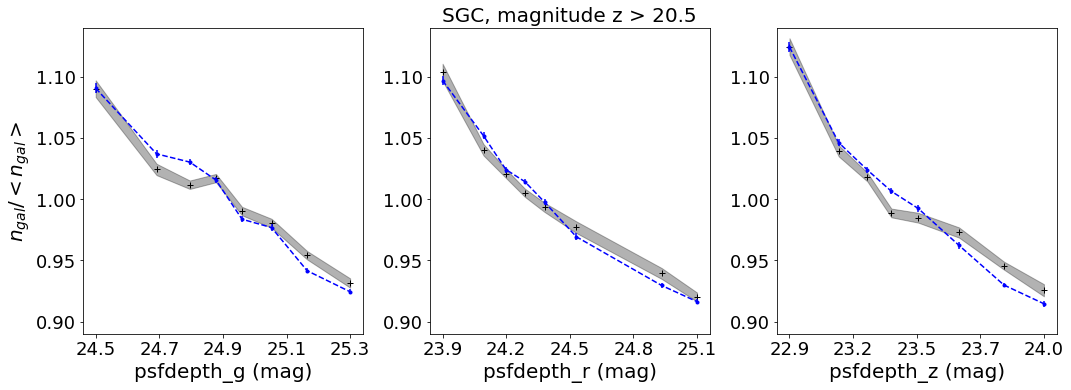}
%\end{subfigure}
    \caption{Systematics trends with \texttt{psfdepth}, when LRGs are split into a bright sample (rows 1\&3) and a faint sample (rows 2\&4), in the NGC (rows 1\&2) and SGC (rows 3\&4). The grey line represents \texttt{DESI-LRGs}, while the blue dashed lines represent \texttt{Obiwan LRGs}. The y-axis in all plots are set to the same range. The result here suggests that imaging systematics for \texttt{DESI LRGs} have a strong depency with their intrinsic magnitude, and this variation is well predicated by \texttt{Obiwan LRGs}.}
    \label{fig:splits}
\end{figure*}

So far, we have only looked at trends between the full LRG target sample and the survey property maps. As the strongest trends we predict are with the imaging depth, one could reasonably expect the significance of the trends to depend on the brightness of the LRG target.
%assumed that imaging systematics is only associated with survey property maps, this is not entirely true. It also correlates with the brightness of the galaxy. 
%To demonstrate the difference
To investigate this, we split the \texttt{DESI LRGs} sample into two sub-samples: the bright sample ($z$ band magnitude $<20.5$), and the faint sample ($z$ band magnitude $>20.5$). With this split, we observe dramatically different trends, as shown in Figure \ref{fig:splits}.

For the brighter half of the data, we find no significant trend observed or predicted with the depth. Conversely, a quite strong trend is observed in the faint half of the data, with $\sim$15\% variation in number density as a function of depth. The trends are well-predicted by \texttt{Obiwan}, with similar performance in both the NGC and SGC. This suggests that a portion of the mismatch found for the SGC in Figure \ref{fig:systematics-SGC} was due to the excess of faint galaxies in the \texttt{Obiwan} sample predicting a somewhat stronger trend in the full sample than observed in the real data.

The results in this section suggest that the best modeling of the variation of the \texttt{DESI LRGs} density with imaging properties would include the brightness of the LRG target. The fiducial DR1 LRG corrections for imaging systematics only split the sample by redshift \cite{DESI2024.II.KP3}. We recommend that the analysis of the sample eventually used for primordial non-Gaussianity, $f_{\rm NL}$, (results of which are to be included in \cite{DESI2024.VIII.KP7C}) makes sure the results are robust to selections on the LRG brightness, as $f_{\rm NL}$ analyses are particularly sensitive to spurious large-scale clustering induced by imaging systematics \cite{rezaie2023local,rezaie2021primordial}.

The \texttt{Obiwan LRGs} predict the brightness-dependent trends with imaging systematics. However, as is shown in Figure \ref{fig:z-deep-dr9}, it is difficult to achieve a perfect match in magnitude distribution between \texttt{DESI LRGs} and \texttt{Obiwan LRGs}. To mitigate this issue, we re-weight \texttt{Obiwan LRGs} so that their magnitude distribution matches with \texttt{DESI LRGs}. We use the basic bin-based reweighting method\footnote{\url{https://hsf-training.github.io/analysis-essentials/advanced-python/45DemoReweighting.html}} to match the magnitude distribution of $g$, $r$, $z$ , $g-r$, and $r-z$ of \texttt{Obiwan LRGs} to \texttt{DESI LRGs}. The matching is successful in that the KS values obtained from the comparison of the \texttt{Obiwan LRGs} and \texttt{DESI LRGs} magnitude distributions are reduced from 0.018, 0.028, 0.038, 0.027, 0.016 to 0.008, 0.008, 0.007, 0.007, 0.006, respectively. We apply these weights to \texttt{Obiwan LRGs} and re-compute the systematics plot with these weights applied. 

\begin{figure}
\includegraphics[width=0.98\linewidth]{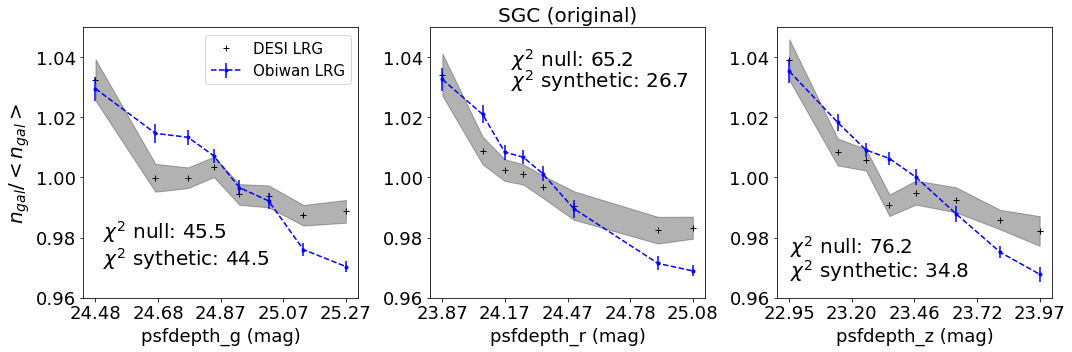}
\hfill
\includegraphics[width=0.98\linewidth]{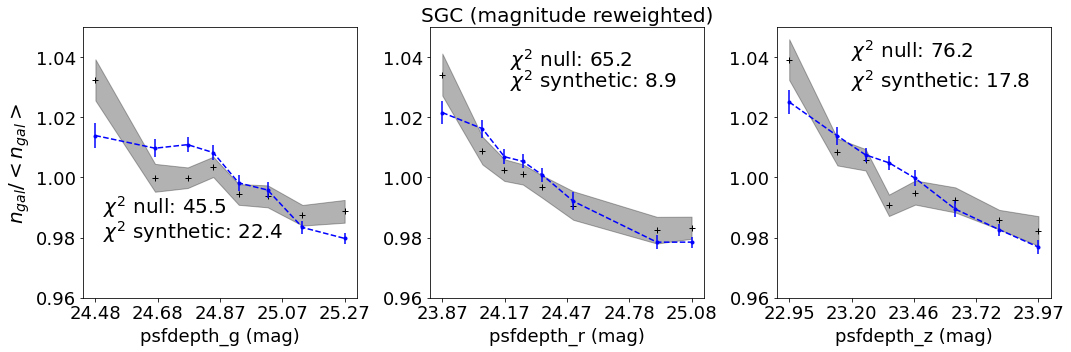}
\caption{The trends in the SGC between \texttt{DESI LRG} and \texttt{Obiwan LRG} density and imaging depth before (upper row) and after (lower row) we apply magnitude re-weighted weights to \texttt{Obiwan LRGs}}
\label{fig:mag-reweight}
\end{figure}

Figure \ref{fig:mag-reweight} shows the trends between LRG density and the \texttt{psfdepth} in the SGC, after re-weighting the \texttt{Obiwan LRGs}. Compared to the result in Figure \ref{fig:systematics-SGC} (We repeated relevant plots in Figure \ref{fig:mag-reweight}), the $\chi^2$ value in \texttt{Obiwan LRGs} is greatly reduced. This re-weighting process is a rudimentary study of how to improve the matching between real and synthetic sources. In regions of different survey properties, the magnitude distribution varies. This method ignores this variation. In addition, although the re-weighted distribution has a much smaller KS value, it is still not small enough to mathematically declare the distributions of \texttt{Obiwan LRGs} and \texttt{DESI LRGs} are drawn from the same underlying distribution. In our future work, we will address this issue thoroughly.  

\subsection{Galactic Extinction}
\label{sec:$E(B-V)$}
Various maps have been developed to correct for galactic extinction. As detailed in \cite{chiang2019extragalactic}, these maps can be divided into three categories: Thermal emission-based maps, e.g., \cite{schlegel1998maps}, stellar reddening-based maps, e.g., \cite{mudur2023stellar, green2018galactic}, and HI-based maps, e.g., \cite{lenz2017new}. The Galactic extinction is typically mapped as $E(B-V)$, in magnitudes, and a correction to apply to any measured magnitude is a constant coefficient \cite{schlafly2011measuring} for the particular wavelength band, multiplied by the $E(B-V)$ at the given celestial coordinate. The SFD (\cite{schlegel1998maps}) map is the most commonly used map for the application of extinction corrections in current galaxy surveys, including the \texttt{DESI LRGs} targets. It is thermal-based, and the data are directly collected from dust emission. This gives it high accuracy. However, it suffers from extragalactic imprints. When stacking the SFD $E(B-V)$ value around reference objects, \cite{chiang2019extragalactic} see that the average $E(B-V)$ value correlates with the redshift of these objects, indicating leaks of Cosmic Infrared Background (CIB) signal. The stellar-reddening-based maps measure dust emission on a different wavelength range, in which CIB contamination is negligible. These maps measure $E(B-V)$ by comparing stellar spectra to templates while fitting for the $E(B-V)$ value at the same location. The accuracy of this method is limited by the number of usable stars at any given location. Moreover, the zero-point calibration uncertainties from star catalogs used for stellar-reddening-based $E(B-V)$ maps could bias the measurements for these maps \cite{rykoff2023dark}. However, they are free from CIB leaks. HI-based maps trace the hydrogen column density, correlated with $E(B-V)$. HI-based maps are not strictly $E(B-V)$ maps. Due to the scope of this work, we did not use HI-based maps in our test.  

Observations of distant galaxies are affected by $E(B-V)$. If we have a ``truth map'' for $E(B-V)$ (and per-band coefficients), it would not be a big problem, as the needed corrections to the photometry would be known. However, we do not have such a ``truth map''. When we detect a correlation of our galaxies' relative density with the $E(B-V)$ map, we are not certain how the error in our $E(B-V)$ map plays a role in such correlation.

We will consider three potential factors that could contribute to the trends we observe between $E(B-V)$ and \texttt{DESI LRGs}:

\begin{enumerate}
    \item \label{item:1} The inaccuracies in the map influence the target selection of \texttt{DESI LRGs} by shifting the dust-corrected flux of each galaxy. 
    
    \item \label{item:2} The $E(B-V)$ values are correlated with the true background galaxy density due to some systematics in the $E(B-V)$ map, e.g., the CIB leaks. For LRGs, their own dust can absorb light and re-emit it in the infrared, contributing to the CIB. Thus, regions with a higher density of LRGs will have higher CIB and a positive trend should be expected between the LRG density and an $E(B-V)$ map with CIB contamination.

    \item \label{item:3} Dust extinction induces more photometric scattering. As galaxies are dimmer at a higher extinction, fewer photons from the galaxy reach the telescope given the same exposure time. This effectively reduces the depth of the image. 
    \texttt{Obiwan} simulates this effect.
\end{enumerate}

Factors~\ref{item:1} and~\ref{item:2} are not recovered by \texttt{Obiwan} because \texttt{Obiwan} assumes no error in the $E(B-V)$ map, while factor~\ref{item:3} should be automatically included during \texttt{Obiwan}'s image simulation procedure. In \texttt{DESI LRGs}, we observe a strong correlation of the relative number density of LRGs with the $E(B-V)$ SFD map, as shown in Figures \ref{fig:systematics-NGC} and \ref{fig:systematics-SGC}. Meanwhile, \texttt{Obiwan LRGs} exhibit a weaker correlation. Under our assumption, the weaker correlation in \texttt{Obiwan LRGs} is due to not simulating factors~\ref{item:1} and~~\ref{item:2} in the \texttt{Obiwan} simulation.

To test the impact of the three factors on the LRG density vs.~$E(B-V)$  trend, we redo the \texttt{DESI LRGs} target selection based on two alternative maps. One of the alternative maps is the stellar-reddening-based map produced by \cite{mudur2023stellar}, which we label `Mudur+23'. This map derives $E(B-V)$ map based on data from Pan-STARRS1 \cite{chambers2016pan} and 2MASS \cite{2006AJ....131.1163S} along with Gaia parallaxes \cite{vallenari2023gaia}. The other alternative map is the Corrected-SFD (CSFD) map produced by \cite{chiang2023corrected}. It cross-correlated the SFD map with the galaxy clustering signal and separated the SFD map into two maps: the CIB residual map and the CSFD map. The CSFD dust map has a much lower residual CIB contamination than SFD. However, the correction is CIB only. Non-CIB systematics are not included in the CSFD map. Compared to the stellar reddening-based map, the $E(B-V)$ values on the CSFD map share more similarity with the SFD map. We define LRGs selected with these maps \texttt{real LRGs}, since they are all real galaxies we observe. When using an alternative $E(B-V)$ map, \texttt{real LRGs} is not always identical to \texttt{DESI LRGs}. When we use the alternative $E(B-V)$ maps, the (extinction-corrected) magnitudes of the galaxies change. When we apply LRG target selection to these galaxies, we obtain a sample slightly different from \texttt{DESI LRGs}. Meanwhile, if we use SFD map in LRG target selection, \texttt{real LRGs} are identical to \texttt{DESI LRGs}.

\begin{figure}[htp] 
    \centering
    \includegraphics[width=0.48\textwidth]{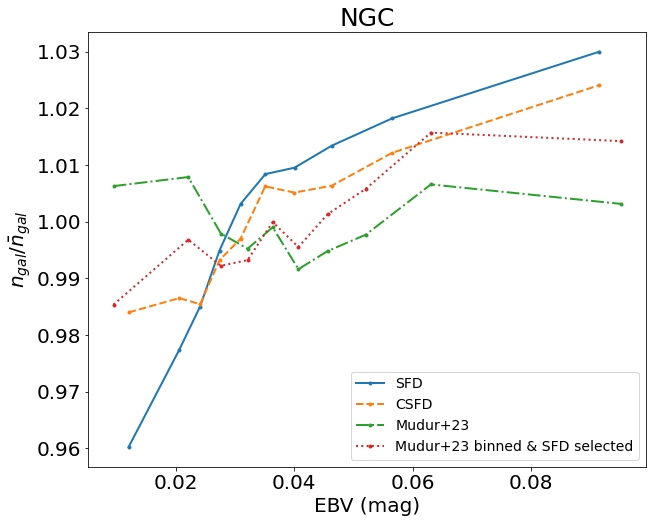}
    \hfill
  \includegraphics[width=0.48\textwidth]{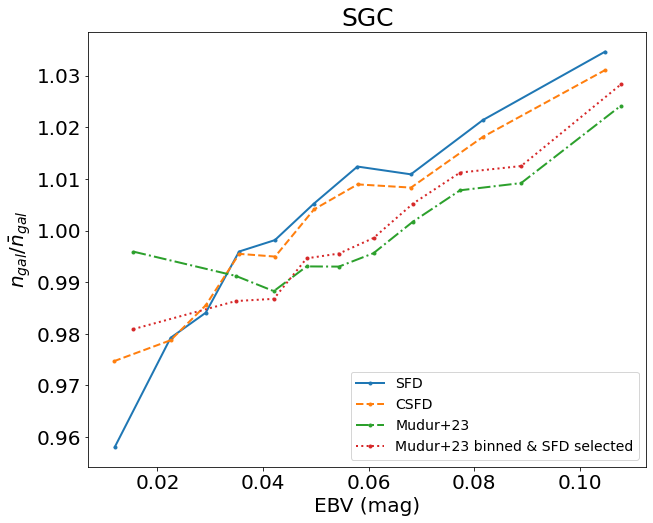}

    \caption{Density fluctuation of \texttt{DESI LRGs} based on different maps in the NGC (left) and SGC (right) region. The target selection process is also altered based on different $E(B-V)$ maps. The blue line is based on the SFD map \cite{schlegel1998maps}, the green line (labeled `Mudur+23') is based on the stellar reddening map from \cite{mudur2023stellar}, the orange line is based on the Corrected-SFD (CSFD) map \cite{chiang2023corrected}, the red line uses LRGs selected by the SFD map, but the binning is computed with the same stellar reddening map as the green line. For LRGs selected and plotted against SFD maps, we see a clear trend. In the stellar reddening map, NGC and SGC are slightly different. In NGC, we do not see any trend. In SGC, we see a trend that is weaker than the SFD map. The difference is also seen in \texttt{Obiwan LRGs} seen in Figure \ref{fig:$E(B-V)$-bright-faint}. In the CSFD map, this trend is similar to the SFD map. In the lower $E(B-V)$ region, the trend is weaker. The fact that the red lines exhibit a larger trend than the green line suggests that some of the $E(B-V)$ systematics trend is introduced by target selection, which uses the $E(B-V)$ to determine the colors of galaxies.}
    \label{fig:$E(B-V)$3maps}
\end{figure}

The trend for the density of the \texttt{real LRGs} against the three $E(B-V)$ maps (the original SFD map and the two alternative maps) are shown as blue, orange, and green curves in Figure \ref{fig:$E(B-V)$3maps}. The result is plotted against the $E(B-V)$ assumed in the target selection (i.e., the $x$-axis is three different quantities). We also show an additional red curve. It uses \texttt{DESI LRGs} selected by the SFD map, but it is plotted against the same stellar-reddening-based map as the green curve (the $x$-axis uses quantities in the stellar-reddening-based map). We define this procedure as ``using a different binning'', as it effectively only changes which $E(B-V)$ bin each LRG goes to. Compared with the blue curve, one can observe that all trends are decreased when using an alternative map for target selection and/or binning. We can use the three factors to explain what we see in Figure \ref{fig:$E(B-V)$3maps}. 

Factor~\ref{item:1} can be seen by comparing the red and green curves. The two samples are binned with the same stellar-reddening-based map. The red sample is \texttt{DESI LRGs} that uses the SFD map to determine their magnitudes and perform target selection. The green curve uses a stellar-reddening-based map for a similar procedure. Thus, the difference between these two curves is a result of the change in target selection. We find that 2.0\% of LRGs are selected with the stellar-reddening based map, and not selected by the SFD map. 2.6\% of LRGs are selected with the SFD map, but not selected by the stellar-reddening based map. 

These two curves in Figure~\ref{fig:$E(B-V)$3maps} show how LRG sample difference results in different systematics trends. The red curve has a larger trend than the green curve, which indicates that the CIB leaks affect the target selection of \texttt{DESI LRGs}. 

Factor~\ref{item:2} can be seen by comparing the blue and red curves. The two trends are computed with identical \texttt{DESI LRGs} sample. The only difference lies in the binning. The blue curve is SFD-map-selected LRGs binned with the SFD map. It has a stronger trend. The red curve is the same SFD-map-selected LRGs but binned with the stellar-reddening-based map. It has a weaker trend. The reduction of trend in the red curve compared to the blue curve is partly a result of less CIB contamination in the stellar-reddening-based map. However, it is worth noting that this trend reduction could also result from uncertainties in the Mudur+23 map: If we add some noise to the ``$E(B-V)$ truth map'', then some $E(B-V)$ values will be falsely assigned to other bins, resulting in a suppression of the actual trend. It could also result from SFD $E(B-V)$ map uncertainty: as the galaxies are selected with the SFD map, binning with the same map includes relevant unknown systematics.  

This contamination within the SFD $E(B-V)$ map has the potential to impact measurements of galaxy clustering. Given that the CIB emanates from galaxies, it exhibits a correlation with galaxy clustering. Thus, the utilization of the SFD $E(B-V)$ map for imaging systematics correction would inadvertently null part of the true clustering signal, and lead to biased clustering measurements. Indeed, based on these considerations, the DESI large-scale structure (LSS) catalogs \cite{DESI2024.II.KP3} in data release 1 (DR1 \cite{DESI2024.I.DR1}) do not directly use the SFD or CSFD maps for the correction of systematic trends.

Factor \ref{item:3} is predicted by \texttt{Obiwan}, since \texttt{Obiwan} simulates the dimming effect from dust extinction. As we discussed in Section \ref{sec:correlation-with-galaxy-brigntness}, galaxies with different apparent magnitudes are affected by photometric scattering differently. Faint galaxies are more sensitive towards changes in the depth of images. Here $E(B-V)$ can generate an effect similar to depth by making galaxies fainter. Given its similarity with depth, we cannot see an isolated effect for Factor \ref{item:3} on \texttt{DESI LRGs}. However, with \texttt{Obiwan} image simulations, we know that these effects are simulated as a whole. If we look at the $E(B-V)$ systematics trend in Figures \ref{fig:systematics-NGC} and \ref{fig:systematics-SGC}, we see that the difference in systematics trend of \texttt{Obiwan LRGs} and \texttt{DESI LRGs} are larger in the NGC than the SGC, which is evidenced by $\chi^2$ synthetic. This value is 100.5 in NGC and 27.7 in SGC. This indicates that a higher proportion of the $E(B-V)$ trend in the SGC is contributed by phenomena simulatable by \texttt{Obiwan}. Factor \ref{item:3} is one of these phenomena. 

\begin{figure*}
    \centering
    \includegraphics[width=.45\linewidth]{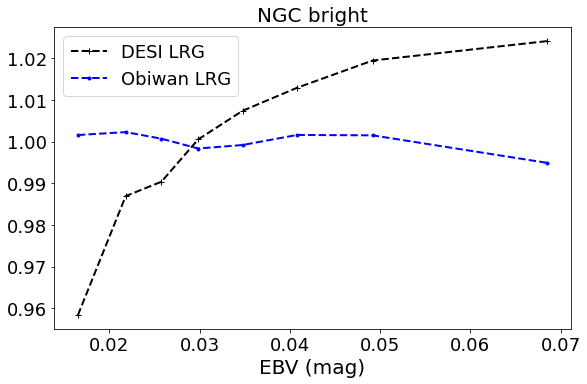}
    \centering
    \hfill
  \includegraphics[width=.45\linewidth]{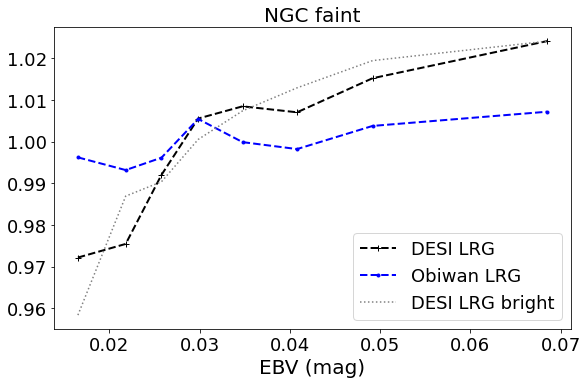}
   \includegraphics[width=.45\linewidth]{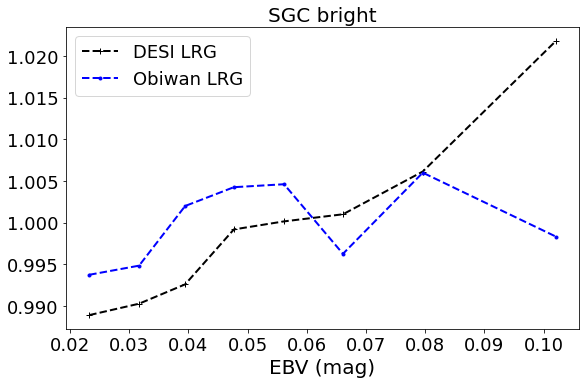}
    \centering
    \hfill
  \includegraphics[width=.45\linewidth]{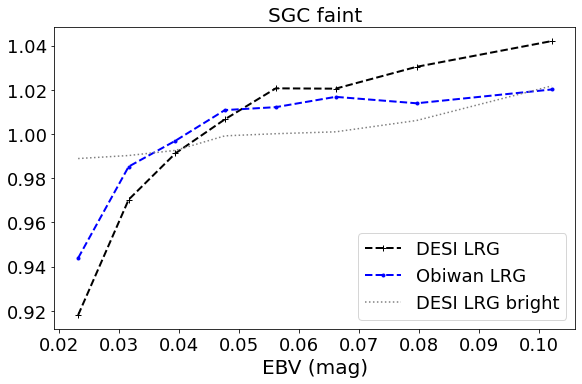}
    \caption{Systematics trend in $E(B-V)$ when we separate the sample into two subsamples: One bright sample with z band magnitude < 20.5, One faint sample with z band magnitude > 20.5. Blue curves are made with Obiwan-LRGs, and black curves are made with real LRGs. The grey curve on the left plots is the same black curve on the right plots, and they are plotted for a better visual comparison. We see some trend recovery in the SGC faint sample from Obiwan-LRGs, meaning part of the trend is due to simulatable phenomena by \texttt{Obiwan} (e.g., $E(B-V)$ in the SGC produces a shallower depth that is not corrected in the observing strategy design). The rest do not match well, indicating that the trend in real LRGs is likely due to the SFD $E(B-V)$ map systematics. The result here supports the observation in Figure \ref{fig:$E(B-V)$3maps}, where we see more discrepancies between the SFD $E(B-V)$ and alternative maps in NGC than in SGC. }
    \label{fig:$E(B-V)$-bright-faint}
\end{figure*}

We take a deeper look at these simulatable phenomena. We noticed that $E(B-V)$ in the SGC is negatively correlated with \texttt{psfdepth} with a correlation coefficient of about $-0.24$. While in NGC it is positively correlated with a correlation coefficient of about $0.14$. The difference in the two regions is due to the survey design strategy. For the DECaLS data, the DECaLS team adjusted the exposure times to try to reach uniform depth after dust extinction \cite{SurveyOps.Schlafly.2023}. The reported \texttt{psfdepth} are in observed magnitudes. Thus, the survey goes to deeper \texttt{psfdepth} where the E(B-V) is larger. The positive correlation is expected in the DECaLS area. The SGC has more DES data, and DES did not adopt the depths in the same way as DECaLS. As lower \texttt{psfdepth} leads to higher relative LRG density, this partially explains why we see a stronger positive trend in \texttt{Obiwan LRGs} in the SGC than in the NGC. Due to its close relation with depth, this trend is also related to the magnitude of LRGs. In Figure \ref{fig:$E(B-V)$-bright-faint}, we split the sample into faint LRGs and bright LRGs in the same way as in Section \ref{sec:correlation-with-galaxy-brigntness}. We see some trend in the SGC faint sample in \texttt{Obiwan LRGs}, while the trend is much weaker in the SGC bright sample. Since faint ones are more influenced by photometric scattering, this behavior is expected. In NGC, given the positive correlation between \texttt{psfdepth} and $E(B-V)$, which cancels out with each other, the trend is much weaker. We can compare the $E(B-V)$ trend of bright \texttt{DESI LRGs} and faint \texttt{DESI LRGs} by looking at the grey and black curve on the left side of Figure \ref{fig:$E(B-V)$-bright-faint}. We see that the difference is larger in the SGC than in the NGC. These all indicate that compared to the NGC, the $E(B-V)$ systematics trend in the SGC has a higher portion contributed by \texttt{Obiwan} simulatable phenomena that could be magnitude dependent. As the overall amplitude of the SFD $E(B-V)$ trend is similar in the NGC and the SGC, this suggests that the contributions of factors \ref{item:1} and \ref{item:2} (unknown $E(B-V)$ systematics, including the persistence of CIB contamination) are larger in the NGC than in the SGC. This result is consistent with the stellar-reddening-based map in Figure \ref{fig:$E(B-V)$3maps}. The stellar-reddening-map-based $E(B-V)$ trend (green) which is devoid of CIB is more similar to the SFD-map-based $E(B-V)$ trend (blue) from the SFD map in the SGC than in the NGC. Our findings here are consistent with \cite{schlafly2010blue} (particularly Figure 15 in their paper), and we further discuss it in the next paragraph.

In Figure \ref{fig:$E(B-V)$3maps}, the two alternative maps we use here have different trends, as shown in the green and orange curves. The CSFD map only corrects for the contributions of CIB in the SFD map, so any other systematics in the SFD map is not recovered by CSFD. As discussed in \cite{mudur2023stellar}, there is a negative trend for star-reddening-based $E(B-V)$ - $E(B-V)$ SFD as a function of $E(B-V)$ SFD. This trend supports the conclusion in \cite{schlafly2010blue}. In their work, they computed the reddening coefficient with the blue tip of the stellar locus and the measurement is based on the $E(B-V)$ SFD map. They found a dust map normalization difference of 15\% between NGC and SGC that may be due to uncertainties in dust temperature correction. The dust temperature map has high uncertainty, particularly in the low $E(B-V)$ region. The angular resolution of the dust temperature map is much lower than the $E(B-V)$ map. As the $E(B-V)$ scales linearly with the dust temperature map, uncertainties in the dust temperature map bias the overall amplitude of $E(B-V)$ rather than the relative fluctuations. This effect generates a trend as a function of the absolute $E(B-V)$ value.

\subsection{Investigation of Trends With \texttt{psfsize}}
\label{sec:psfsize}
\texttt{Obiwan LRGs} predict that the LRG target density generally increases with \texttt{psfsize}, and if anything, the opposite seen in \texttt{DESI LRGs} (though the trends are consistent with the null expectation). To locate the problem, we first split the sample into \texttt{faint LRGs} and \texttt{Bright LRGs} in the same way as in Section \ref{sec:correlation-with-galaxy-brigntness}. The discrepancy is mainly contributed by \texttt{faint LRGs}. We further split \texttt{faint LRGs} into a high-depth region with \texttt{psfdepth\_z} $> 23.5$ and a low-depth region with \texttt{psfdepth\_z} $< 23.5$. We find that the trend is mainly contributed from the low-depth region.

\begin{figure}
\centering
\includegraphics[width=1.\columnwidth]{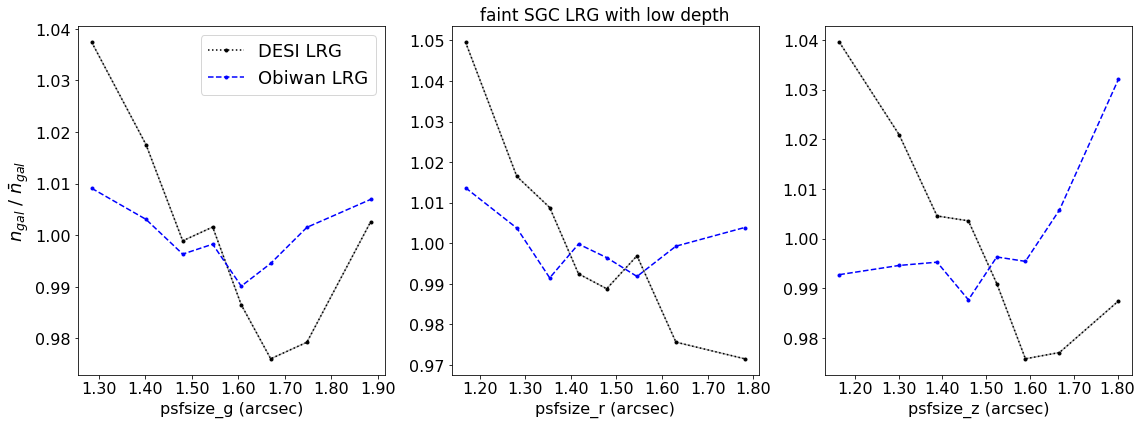} 
\caption{The \texttt{DESI LRG} \texttt{psfsize} systematics trend. The sample is in the SGC and has a magnitude cut of magnitude $z > 20.5$, and a depth cut of \texttt{psfdepth z} $ <23.5$. We perform these cuts because the trend is most evident in faint LRGs in low-depth regions. The black dashed line is the result from \texttt{DESI LRGs}, and the blue dashed line is the trend from \texttt{Obiwan LRGs}.}
\label{fig:psfsize_z_mismatch}
\end{figure}

\begin{figure}
    \centering
        \includegraphics[width=0.48\textwidth]{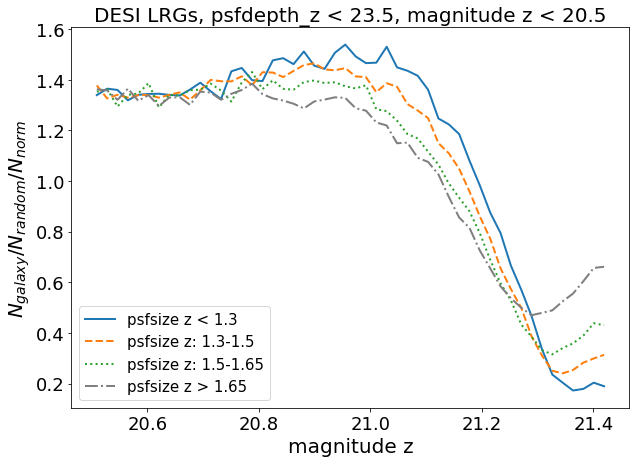}
    \hfill
        \includegraphics[width=0.48\textwidth]{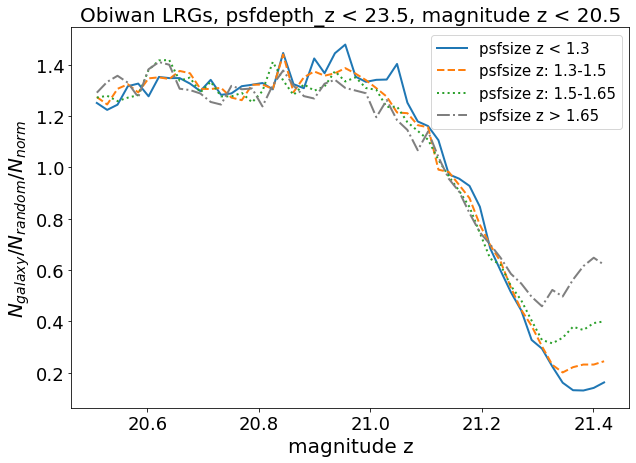}
    \caption{Histogram of $z$ band magnitude of \texttt{DESI LRGs} (left) and \texttt{Obiwan LRGs} (right) in the SGC. The sample is split into 4 sets with different \texttt{psfsize\_z} range. The histogram is normalized with the randoms in the same footprint so that the amplitude reflects their relative density.}
    \label{fig:psfsize_splits}
\end{figure}

 Figure \ref{fig:psfsize_z_mismatch} shows the trend of \texttt{DESI LRGs} and \texttt{Obiwan LRGs} with these cuts (magnitude $z$ > 20.5, \texttt{psfdepth\_z} < 23.5). They are composed of 19.5\% of all LRGs in the SGC. The trend of \texttt{DESI LRGs} goes down first. And then it goes up for \texttt{psfsize\_g} and \texttt{psfsize\_z}. However, \texttt{Obiwan LRGs} does not have a noticeable downward trend, while it does have an upward trend at large \texttt{psfsize}. Plots in Figure \ref{fig:psfsize_splits} indicate the reason behind the mismatch. Here, we split the sample into four regions by \texttt{psfsize\_z}, then plot their histograms, binning in $z$ magnitude. We normalize these histograms with the random galaxies in the same regions. Using this procedure, we can compare the relative densities of each region as a function of the $z$ band magnitude. The plot for \texttt{Obiwan LRGs} is produced in the same way as \texttt{DESI LRGs}. The variation we see in \texttt{DESI LRGs} around $z$ band magnitude 21 is not replicated in \texttt{Obiwan}. 

\begin{figure}
\centering
\includegraphics[width=0.7\linewidth]{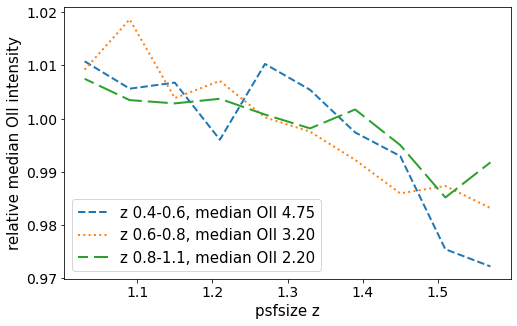}
\caption{The median flux of OII emission line for LRGs at three redshift bins in different \texttt{psfsize\_z} range, divided by the overall median OII flux. Blue is for the redshift range 0.4-0.6, orange is for the redshift range 0.6-0.8, and green is for the redshift range 0.8-1.1. The trend here originates from a higher loss in flux for \texttt{disk LRGs} at high \texttt{psfsize\_z}}
\label{fig:OII}
\end{figure}

The COSMOS repeat sets are ideal sets for pinning down this issue. We perform several tests documented in Appendix \ref{appendix:psfsize}. We eventually concluded that this effect is caused by a correlation of flux bias with \texttt{psfsize}, shown in Figure \ref{fig:cosmos_repeats_psfsize_trend}. The trend for this bias is larger for galaxies with \sersic\, index $>=$4 (we define them as \texttt{disk LRGs}). Because of this dependency with galaxy morphology, \texttt{disk LRGs} have a lower flux at high \texttt{psfsize\_z}, so they are less likely to be selected as LRGs due to the color cut \ref{equ:fibermagz}. Consequently, we see less \texttt{disk LRGs} in regions with large \texttt{psfsize}. This effect is further propagated to DESI LRGs after spectroscopic redshift cuts. As disk LRGs have a star-forming disk, they have stronger OII emission lines. We use the DESI LRGs with spectroscopic redshift from the \texttt{iron} catalog. We compute the median OII flux for each \texttt{psfsize\_z} bin. This gives us an estimate of the median intensity of OII flux for each \texttt{psfsize\_z} bin. In Figure \ref{fig:OII}, we found noticeable trends in all three redshift bins. The median OII intensity decreases with an increase in \texttt{psfsize\_z}. We do not see similar trends in the north (BASS/MzLS region). As we have not thoroughly tested the north in this work, we do not know why similar trends are not seen in the north. 

 \subsubsection{Possible root causes}
 \label{sec:possible-causes}
Knowing where the discrepancy originates will help us develop treatments for this issue, so we did some further investigation. Our main suspects of root causes are background subtraction and PSF uncertainty. 

We currently use the original background determination by \textsc{Legacypipe} as truth. \texttt{Obiwan} injects galaxies on images that are already background subtracted. Possible residuals or over-subtractions in the background subtraction are not replicated in \texttt{Obiwan}. The background subtraction procedure in the \textsc{Legacypipe} code masks out regions containing sources and fits a spline model \cite{blanton2011improved} to the rest of the images. 

In the case of PSF uncertainty, it is a widely studied subject. In Dark Energy Survey (DES) analysis \cite{zuntz2018dark} \cite{jarvis2016science}, people saw a bias between the PSF size and the size of stars (Brighter-Fatter effect). They also saw a color dependency in the fitting residuals (Differential Chromatic Refraction). The pipeline (PSFEx\cite{bertin2011automated}) used for PSF determination in \texttt{Legacysurvey} is similar to DES (PIFF\cite{jarvis2021dark}), so similar systematics could also be present in \texttt{Legacysurvey}.

We design a test to separate the impacts of background subtraction uncertainty and PSF uncertainty. We apply a Gaussian filter to each individual image (CCD) to make them reach the same \texttt{psfsize}. Then, we measure the 7" aperture flux of the same LRGs on different CCDs. A 7" aperture is large enough to contain most of the flux from LRGs. The 7" aperture flux is independent of any fitting methods or determination of PSF. On each CCD, we record the measured flux and the \texttt{psfsize} value before Gaussian filtering. These data are later used to test the correlation between the measured flux and the \texttt{psfsize}. We perform the same test with synthetic \texttt{Obiwan LRGs} on real images and also on fully synthetic images with Gaussian background noise for comparison.

We select all pairs of images with a difference in \texttt{psfsize\_z} > 0.12". With these data pairs, we found that the 7" aperture flux for the images with a larger PSF is 0.19 nanomaggies lower than the images with a smaller PSF. The equivalent value measured from the synthetic \texttt{Obiwan LRGs} with synthetic images is less than $\sim$10$^{-3}$ nanomaggies. For \texttt{Obiwan LRGs} injected on real images, this number is 0.09 nanomaggies. Overall, we conclude that the background on a larger \texttt{psfsize} is lower than that on images with smaller \texttt{psfsize}. The flux loss in \texttt{DESI LRGs} is $\sim$0.1 nanomaggies larger than the synthetic \texttt{Obiwan LRGs}. However, we do not find the flux loss for \texttt{Disk LRGs} different from \texttt{Compact LRGs}. The conclusion here applies to 3 optical ($g$, $r$ $z$) bands.

Our test on 7" aperture flux suggests that there could be some flux bias induced by background subtraction. In regions with larger \texttt{psfsize}, it becomes more difficult to distinguish between signals from real galaxies and signals from the background, as the galaxies on images are more spread out. The signal-to-noise is lower on each pixel. The procedure of source fitting uses a set of effective pixels. The fitting of \texttt{Disk LRGs} requires more pixels, so it is more sensitive to the cumulative effect of the flux bias from the background subtraction.

On the other hand, uncertainties from PSF determination may not be a dominant factor here, as our PSF-insensitive test is already sufficient to explain the trends with \texttt{psfsize}.

\subsection{Other secondary effects}
\label{sec:other_effects}
We have identified other effects that would potentially affect our ability to properly simulate the expected target density fluctuations in the real data. We do not see an evident mismatch when these effects are not being accounted for. However, we list these effects and potential treatments below. 

Shape noise: Real galaxies are not perfect model galaxies. Since our injected galaxies \emph{are} accurately described by the simple galaxy models, they would be slightly different from what we see in real galaxies. This effect is negligible for the precision we are worried about now. However, this would be a more significant issue when trying to study galaxies' shape response, e.g., as done for weak lensing measurements \cite{abbott2022dark}, and the studies of magnification \cite{elvin2022dark}. Realistic galaxy stamps will be needed in this analysis. 

Sagittarius stream \cite{newberg2002ghost}: No noticeable trend due to the Sagittarius stream being present in the LRGs. The contamination level from stars is extremely low for LRGs, so stars do not provide direct contamination. However, the Sagittarius stream, along with other stellar streams or Milky Way stars, can enter contamination by altering the background level of the image. As stellar streams originate from satellite galaxies, they contain many faint stars below the detection limit and merge into the background. We may get a shifted flux measurement when falsely subtracting background noise that is faint Sagittarius stream stars. 

Variation in extinction coefficients \cite{schlafly2011measuring}: The extinction coefficient denotes the extent of extinction's impact on each band. It is computed with stellar spectra. However, this coefficient is related to the wavelength of observation. This means that if the galaxy we observed differs from stellar spectra, the coefficient we assume would not be accurate.

\subsection{Discussion on imaging systematics mitigation}
\label{sec:imgsys-discuss}
Currently, studies on imaging systematics mitigation typically employ an approach based on survey property maps. The observed footprint for a survey is separated into small pixels. Each pixel is associated with a collection of quantities like depth, PSF, $E(B-V)$, airmass, etc. Thus, we have pixelized maps for each of these quantities. Studies on imaging systematics mitigation typically find a relationship between these survey property maps and the false fluctuations of galaxy densities. Then, these studies derive ``imaging systematics weights''. The weights minimize the systematics trend on all these survey property maps. One weight is assigned to one pixel so that the weights form an imaging-systematics-weight map with the same pixel resolution. These kinds of methods rely on two assumptions that are not entirely true:

\begin{itemize}
    \item \textbf{Assumption 1:} No errors exist in these survey property maps.
    \item \textbf{Assumption 2:} For sources at the same location, the imaging systematics associated with them are the same and can be represented as a number that we call ``imaging systematics weight''. 
\end{itemize}

Our study with galactic extinction shows flaws on \textbf{Assumption 1}. The SFD $E(B-V)$ map is commonly used as a survey property map. Therefore, such methods have the assumption that the SFD $E(B-V)$ map is perfectly accurate. After applying systematics weight, the sample tends to barely have any trend against the SFD $E(B-V)$ map. Our study challenges this approach of treating the SFD $E(B-V)$ map. Our findings imply that a portion of the observed trend within the SFD $E(B-V)$ map may be attributable to contamination stemming from CIB or other Large-Scale-Structure systematics. In Section \ref{sec:$E(B-V)$}, factor~\ref{item:1} could result in false fluctuations of galaxies that need to be corrected with weights. Since target selection defines color cut to select LRGs, this shift is most evident around the color selection boundaries. This effect needs to be carefully corrected since its impact on different magnitudes of LRGs may vary significantly. Factor~\ref{item:2} is purely due to false binning and should not be corrected at all. Ideally, systematics-weighted galaxies should still have a positive trend against the SFD $E(B-V)$ map, as we should not correct the trend that stems from factor~\ref{item:2}. 

Our study on galaxy brightness in Section \ref{sec:correlation-with-galaxy-brigntness} and \texttt{psfsize} in Section \ref{sec:psfsize} challenges \textbf{Assumption 2}. We show that the imaging systematics trend for DESI LRGs changes significantly when we apply a cut on $z$-band magnitude. The mechanism behind this is well-understood: Faint galaxies have more photometric scattering than bright galaxies so they are affected by the variation of galaxy depth and seeing (\texttt{psfsize}) to a larger extent. If two galaxies of different magnitudes have the same configuration in a survey property map, they could be affected by imaging systematics differently. However, the current survey-property-map-based approach will assign them the same ``imaging systematics weight''. Applying the same correction for galaxies with different brightness will result in an over-correction in brighter galaxies and an under-correction in fainter galaxies. In both cases, they introduce an artificial signal when computing the sample's power spectrum. 

One way to mitigate this issue is to separate galaxy samples into subsamples with different magnitude ranges. Then, we compute imaging systematics weights independently for each subsample. However, with a lower density for each subsample, the computation of imaging systematics weight would be prone to fitting the noise of each subsample. Thus, such a step needs to be carefully validated. 

Another way to deal with the issue is through forward modeling tools like \texttt{Obiwan}. Forward modeling can produce randoms with detailed scattering features for each source at any location. We can use such randoms to correct for magnitude-dependent imaging systematics. The randoms produced from forward modeling can predict density fluctuations at any color and brightness. Compared with making such predictions with real data, there are many advantages of predicting this density fluctuation with forward modeling randoms like \texttt{Obiwan LRGs}. These randoms do not contain clustering signals, ensuring recovery of imaging systematics trends solely from uniform random fluctuations. They also lack cosmic variance: Their covariance with survey property maps is smaller than real galaxies under the assumption of the same number density. Moreover, they can reach a number density larger than real galaxies with more simulations. These advantages would help make the predictions precise and free from over-fitting the true clustering signal. 

Due to the scope of this work, we did not perform clustering measurements. Current \texttt{Obiwan LRGs} does not have a large enough density to be used as magnitude-dependent randoms. In Appendix \ref{appendix:speedup}, we explore various methods to enhance the computational efficiency of generating \texttt{Obiwan} randoms. 

\subsection{Implications for cosmological analysis}
\label{sec:cosmology-analysis}
For the robustness of cosmological analyses, accurately estimating and mitigating the effects of imaging systematics is crucial and will become increasingly challenging as statistical uncertainties decrease. Our work has discussed several phenomena that may go under the radar of many systematics mitigation methods: A sample could have a null systematics trend over a surveyed property in terms of number density, but its redshift distribution, galaxy bias, etc., could still change. \cite{dominguez2023galaxies} shows that galaxies' morphology correlates with the Large-Scale-Structure environment they stay in, indicating that the morphology dependence possibly links to bias differences. The changes in the galaxy properties and thus (likely) galaxy bias parameters may complicate measurements of structure growth that combine galaxy auto correlations and galaxy-lensing cross-correlations (e.g., \cite{white2022cosmological}), as these typically assume the same effective galaxy bias parameters in the modeling to recover the amplitude of matter clustering. The variation in galaxy bias is particularly problematic for primordial non-Gaussianity measurements. Many survey properties vary considerably on large scales (e.g., $E(B-V)$). When the galaxy bias shifts with these survey properties, it effectively introduces a scale-dependent bias that could degenerate with the primordial non-Gaussianity signal. \cite{rezaie2024local} found that the need to include many maps in the modeling of imaging systematics inflated the statistical uncertainty by more than a factor of two. By correctly modeling how the intrinsic properties of galaxies change with survey property maps, we can reduce this statistical uncertainty induced by imaging systematics. 

\section{Conclusions}
We conclude our findings as follows:
\begin{itemize}
    \item To properly simulate the \texttt{DESI LRGs} sample, we further develop the pipeline, \texttt{Obiwan} to be compatible with both optical band images and the infrared band WISE images. 
    \item We find shape bias in the COSMOS Deep catalog. We develop a procedure to eliminate this shape bias successfully, see Section \ref{sec:truth-data-generation} and Appendix \ref{appendix:de-bias}. 
    \item Our inclusion of WISE images demonstrates how forward modeling can be used for combined survey analysis. We find significant differences in the WISE flux measured as a function of the imaging depth (see Figure \ref{fig:depth_compare}), due to the increased number of total sources that the WISE flux can be attributed to in the deeper data. In the limited efforts of this work, we did not validate the optimal optical band depth to be used in the WISE flux measurement. However, it is possible to perform this cross-survey validation with forward modeling tools like \texttt{Obiwan}. For example, LSST can be combined with Euclid and Roman telescope to reach a broader color space. Such a combination may encounter similar issues, as the depth limit is different across surveys.

   \item We find a consistent negative bias in flux measurement in all bands. This bias is positive in synthetic images with artificial blending and Gaussian noise. Indicating that this bias is likely due to the background noise structures in the real images. 
    \item We see imprints of systematics in extinction maps based on dust emission (e.g. CIB contamination, variation in dust properties that changes the optical-infrared relation). We find significant trends between the \texttt{DESI LRGs} density on the sky and the SFD $E(B-V)$ map \cite{schlegel1998maps}, which are not predicted by \texttt{Obiwan}. These trends change significantly when we test against alternative $E(B-V)$ maps and are nearly removed when compared to the stellar-reddening-based $E(B-V)$ map \cite{mudur2023stellar}. Since CIB is correlated with galaxy clustering, we caution against any analyses of \texttt{DESI LRGs} eliminating this trend, as this would result in removing real large-scale structures. Meanwhile, some extent of the trends with $E(B-V)$ is real and related to the impact on the imaging depth (photometric scattering), and is predicted by our forward modeling. 
    \item We find that \texttt{Obiwan} predicts the trends with \texttt{psfdepth} that we observe in the real data, which are similarly strong in each of the optical imaging bands. 
    While the overall trends are predicted, statistically, they are not consistent in the SGC. The slight mismatch we see is due to cosmic variance in the truth input, which has a larger impact on the deepest data in the SGC. The mismatch in the color and magnitude distribution of \texttt{Obiwan LRGs} and \texttt{DESI LRGs} is the main source of the systematics trend deviation in this region. We further demonstrate that this issue can be greatly mitigated by applying a  `color matching' weight to \texttt{Obiwan LRGs} to match its color distribution with that of \texttt{DESI LRGs}. The details are discussed in Section \ref{sec:correlation-with-galaxy-brigntness}.
    
    \item We find that the trends with imaging depth depend strongly on the $z$-band magnitude. When selecting the data to have $z<20.5$, only ~1\% level variations are found in the observed data with the imaging depth, and \texttt{Obiwan} predicts no significant trend. However, with selection cut $z>20.5$, up to 20\% trends are observed in the real data and predicted by \texttt{Obiwan}. This suggests it is important to properly model the color/brightness dependence of systematic trends in the \texttt{DESI LRGs} target sample when producing any corrections for use in galaxy clustering measurements. Forward modeling randoms like \texttt{DESI LRGs} can be a powerful catalog to correct for imaging systematics that varies with intrinsic galaxy brightness.
    
    \item We find signs of non-simulated density variation against \texttt{psfsize}, which is especially prominent for extended galaxies. This is likely due to a bias in background subtraction that is correlated with \texttt{psfsize} for each CCD. This effect results in LRGs around color selection boundaries having different properties, and the impact is also seen in LRGs after spectroscopic redshift cut.  
    
    \item We need to know how imaging systematics from each survey property are generated before developing methods to correct for these trends. Current imaging systematics methods commonly take the approach of developing one weight for each \textsc{HealPix} pixel. We demonstrate that this may not be enough if we want to reach a higher precision. Uncertainties in survey property maps, and the magnitude dependence of galaxies on some survey properties, could induce imperfect mitigation in imaging systematics that cannot be recovered with these methods. Signals like primordial non-Gaussianity are highly degenerate with imaging systematics signals. These factors can generate unknown bias in such cosmological measurements. 
    
    \item By comparing systematic trends in real galaxies with those in synthetic galaxies generated by forward modeling tools like \texttt{Obiwan}, we can detect potential unknown imaging systematics and explore relevant resources to identify their root causes. Replicating imaging systematics in real galaxies through forward modeling tools like \texttt{Obiwan} is challenging work. A perfect match between real and synthetic galaxies is only possible when all potential contaminations are properly considered in the image simulation settings. However, a mismatch of density fluctuations in the synthetic galaxies also provides us valuable insights into what unknown systematics could possibly be missed. Identifying missing factors is vital for proper imaging systematics mitigation. 
    
\end{itemize}

Indeed, the results we have described have already been used to help understand the LRG sample used for DESI DR1 \cite{DESI2024.I.DR1,DESI2024.II.KP3} and thus the cosmological analyses \cite{DESI2024.III.KP4,DESI2024.V.KP5,DESI2024.VI.KP7A,DESI2024.VII.KP7B,DESI2024.VIII.KP7C}.

\section*{Data Availability}
We produce the \texttt{Obiwan LRGs} catalog that traces imprints of \texttt{DESI LRGs} in the real images and can be used for future studies. The product is currently available internally for DESI members \footnote{\url{https://data.desi.lbl.gov/desi/survey/catalogs/image\_simulations/LRG/}}. The data will be uploaded to zenodo before publication\footnote{\url{https://zenodo.org/uploads/11260267}}. In the future, the catalogs used in this analysis will be made public along with the Data Release 1 \footnote{ \url{https://data.desi.lbl.gov/doc/releases/}}. 

\section*{Acknowledgements}
We thank Yi-Kuan Chiang for insightful discussions on dust extinction maps.

We thank Alex Krolewski and the anonymous reviewer for reading and providing feedback on this work.

HK thanks the people from the NERSC helpdesk for support in installing and compiling software in NERSC. 

The results lead in this paper have received funds from MCIN/AEI/10.13039/501100011033 and UE NextGenerationEU/PRTR (JDC2022-049551-I).

AP acknowledges support from the European Union’s Horizon Europe program under the Marie Skłodowska-Curie grant agreement 101068581.

This material is based upon work supported by the U.S. Department of Energy (DOE), Office of Science, Office of High-Energy Physics, under Contract No. DE–AC02–05CH11231, and by the National Energy Research Scientific Computing Center, a DOE Office of Science User Facility under the same contract. Additional support for DESI was provided by the U.S. National Science Foundation (NSF), Division of Astronomical Sciences under Contract No. AST-0950945 to the NSF’s National Optical-Infrared Astronomy Research Laboratory; the Science and Technology Facilities Council of the United Kingdom; the Gordon and Betty Moore Foundation; the Heising-Simons Foundation; the French Alternative Energies and Atomic Energy Commission (CEA); the National Council of Humanities, Science and Technology of Mexico (CONAHCYT); the Ministry of Science and Innovation of Spain (MICINN), and by the DESI Member Institutions: \url{https://www.desi.lbl.gov/collaborating-institutions}.

The DESI Legacy Imaging Surveys consist of three individual and complementary projects: the Dark Energy Camera Legacy Survey (DECaLS), the Beijing-Arizona Sky Survey (BASS), and the Mayall z-band Legacy Survey (MzLS). DECaLS, BASS and MzLS together include data obtained, respectively, at the Blanco telescope, Cerro Tololo Inter-American Observatory, NSF’s NOIRLab; the Bok telescope, Steward Observatory, University of Arizona; and the Mayall telescope, Kitt Peak National Observatory, NOIRLab. NOIRLab is operated by the Association of Universities for Research in Astronomy (AURA) under a cooperative agreement with the National Science Foundation. Pipeline processing and analyses of the data were supported by NOIRLab and the Lawrence Berkeley National Laboratory. Legacy Surveys also uses data products from the Near-Earth Object Wide-field Infrared Survey Explorer (NEOWISE), a project of the Jet Propulsion Laboratory/California Institute of Technology, funded by the National Aeronautics and Space Administration. Legacy Surveys was supported by: the Director, Office of Science, Office of High Energy Physics of the U.S. Department of Energy; the National Energy Research Scientific Computing Center, a DOE Office of Science User Facility; the U.S. National Science Foundation, Division of Astronomical Sciences; the National Astronomical Observatories of China, the Chinese Academy of Sciences and the Chinese National Natural Science Foundation. LBNL is managed by the Regents of the University of California under contract to the U.S. Department of Energy. The complete acknowledgments can be found at \url{https://www.legacysurvey.org/}.

Any opinions, findings, and conclusions or recommendations expressed in this material are those of the author(s) and do not necessarily reflect the views of the U. S. National Science Foundation, the U. S. Department of Energy, or any of the listed funding agencies.

The authors are honored to be permitted to conduct scientific research on Iolkam Du’ag (Kitt Peak), a mountain with particular significance to the Tohono O’odham Nation.

\bibliographystyle{JHEP}  
\bibliography{bib, DESI2024}

\section*{Appendix} % Use \section* to avoid numbering
\appendix
\input{appendix.tex}

\end{document}

%% file: authors.tex
\emailAdd{hkong@ifae.es}

\author[1]{{H.~Kong}\orcidlink{0000-0001-8731-1212
},}
\author[2,3,4]{{A.~J.~Ross}\orcidlink{0000-0002-7522-9083},}
\author[2,3,4]{{K.~Honscheid},}
\author[5,6]{{D.~Lang},}
\author[7]{{A.~Porredon}\orcidlink{0000-0002-2762-2024},}
\author[8]{{A.~de~Mattia},}
\author[9]{{M.~Rezaie}\orcidlink{0000-0001-5589-7116},}
\author[10]{{R.~Zhou},\orcidlink{0000-0001-5381-4372}}
\author[10]{{E.~F.~Schlafly},\orcidlink{0000-0002-3569-7421}}
\author[11]{{J.~Moustakas}\orcidlink{0000-0002-2733-4559},}
\author[12]{{A.~Rosado-Marin},}
\author[10]{{J.~Aguilar}\orcidlink{0000-0001-5381-4372},}
\author[13]{{S.~Ahlen}\orcidlink{0000-0001-6098-7247},}
\author[14]{{D.~Brooks},}
\author[10]{{E.~Chaussidon}\orcidlink{0000-0001-8996-4874},}
\author[10]{{T.~Claybaugh},}
\author[15]{{S.~Cole}\orcidlink{0000-0002-5954-7903},}
\author[16]{{A.~de la Macorra}\orcidlink{0000-0002-1769-1640},}
\author[17]{{Arjun~Dey}\orcidlink{0000-0002-4928-4003},}
\author[18]{{Biprateep~Dey}\orcidlink{0000-0002-5665-7912},}
\author[14]{{P.~Doel},}
\author[19,20]{{K.~Fanning}\orcidlink{0000-0003-2371-3356},}
\author[21,22]{{J.~E.~Forero-Romero}\orcidlink{0000-0002-2890-3725},}
\author[23,24,25]{{E.~Gaztañaga},}
\author[10]{{S.~Gontcho A Gontcho}\orcidlink{0000-0003-3142-233X},}
\author[26]{{G.~Gutierrez},}
\author[27]{{C.~Howlett}\orcidlink{0000-0002-1081-9410},}
\author[17]{{S.~Juneau},}
\author[10]{{A.~Kremin}\orcidlink{0000-0001-6356-7424},}
\author[10]{{M.~Landriau}\orcidlink{0000-0003-1838-8528},}
\author[10]{{M.~E.~Levi}\orcidlink{0000-0003-1887-1018},}
\author[1,28]{{M.~Manera}\orcidlink{0000-0003-4962-8934},}
\author[2,3,4]{{P.~Martini}\orcidlink{0000-0002-4279-4182},}
\author[17]{{A.~Meisner}\orcidlink{0000-0002-1125-7384},}
\author[1,28]{{R.~Miquel},}
\author[29]{{E.~Mueller},}
\author[30]{{A.~D.~Myers},}
\author[18]{{J.~ A.~Newman}\orcidlink{0000-0001-8684-2222},}
\author[31]{{J.~Nie}\orcidlink{0000-0001-6590-8122},}
\author[32,33]{{G.~Niz}\orcidlink{0000-0002-1544-8946},}
\author[5,6,34]{{W.~J.~Percival}\orcidlink{0000-0002-0644-5727},}
\author[10,35,36]{{C.~Poppett},}
\author[37]{{F.~Prada}\orcidlink{0000-0001-7145-8674},}
\author[38]{{G.~Rossi},}
\author[39]{{E.~Sanchez}\orcidlink{0000-0002-9646-8198},}
\author[10]{{D.~Schlegel},}
\author[40,41]{{M.~Schubnell},}
\author[12]{{H.~Seo}\orcidlink{0000-0002-6588-3508},}
\author[17]{{D.~Sprayberry},}
\author[40,41]{{G.~Tarl\'{e}}\orcidlink{0000-0003-1704-0781},}
\author[16]{{M.~Vargas-Maga\~na}\orcidlink{0000-0003-3841-1836},}
\author[17]{{B.~A.~Weaver},}
\author[31]{{H.~Zou}\orcidlink{0000-0002-6684-3997},}

\affiliation[1]{Institut de F\'{i}sica d’Altes Energies (IFAE), The Barcelona Institute of Science and Technology, Campus UAB, 08193 Bellaterra Barcelona, Spain}
\affiliation[2]{Center for Cosmology and AstroParticle Physics, The Ohio State University, 191 West Woodruff Avenue, Columbus, OH 43210, USA}
\affiliation[3]{Department of Astronomy, The Ohio State University, 4055 McPherson Laboratory, 140 W 18th Avenue, Columbus, OH}
\affiliation[4]{The Ohio State University, Columbus, 43210 OH, USA}
\affiliation[5]{Perimeter Institute for Theoretical Physics, 31 Caroline St N, Waterloo, ON N2L 2Y5, Canada}
\affiliation[6]{Waterloo Centre for Astrophysics, University of Waterloo, 200 University Ave W, Waterloo, ON N2L 3G1, Canada} 
\affiliation[7]{Ruhr University Bochum, Faculty of Physics and Astronomy, Astronomical Institute, German Centre for Cosmological Lensing, 44780 Bochum, Germany}
\affiliation[8]{IRFU, CEA, Universit\'{e} Paris-Saclay, F-91191 Gif-sur-Yvette, France}
\affiliation[9]{Department of Physics, Kansas State University, 116 Cardwell Hall, Manhattan, KS 66506, USA}
\affiliation[10]{Lawrence Berkeley National Laboratory, One Cyclotron Road, Berkeley, CA 94720, USA}
\affiliation[11]{Department of Physics \& Astronomy, Siena College, 515 Loudon Road, Loudonville, NY, USA 12211}
\affiliation[12]{Department of Physics \& Astronomy, Ohio University, Athens, OH 45701, USA}
\affiliation[13]{Physics Dept., Boston University, 590 Commonwealth Avenue, Boston, MA 02215, USA}
\affiliation[14]{Department of Physics \& Astronomy, University College London, Gower Street, London, WC1E 6BT, UK}
\affiliation[15]{Institute for Computational Cosmology, Department of Physics, Durham University, South Road, Durham DH1 3LE, UK}
\affiliation[16]{Instituto de F\'{\i}sica, Universidad Nacional Aut\'{o}noma de M\'{e}xico,  Cd. de M\'{e}xico  C.P. 04510,  M\'{e}xico}
\affiliation[17]{NSF NOIRLab, 950 N. Cherry Ave., Tucson, AZ 85719, USA}
\affiliation[18]{Department of Physics \& Astronomy and Pittsburgh Particle Physics, Astrophysics, and Cosmology Center (PITT PACC), University of Pittsburgh, 3941 O'Hara Street, Pittsburgh, PA 15260, USA}
\affiliation[19]{Kavli Institute for Particle Astrophysics and Cosmology, Stanford University, Menlo Park, CA 94305, USA}
\affiliation[20]{SLAC National Accelerator Laboratory, Menlo Park, CA 94305, USA}
\affiliation[21]{Departamento de F\'isica, Universidad de los Andes, Cra. 1 No. 18A-10, Edificio Ip, CP 111711, Bogot\'a, Colombia}
\affiliation[22]{Observatorio Astron\'omico, Universidad de los Andes, Cra. 1 No. 18A-10, Edificio H, CP 111711 Bogot\'a, Colombia}
\affiliation[23]{Institut d'Estudis Espacials de Catalunya (IEEC), 08034 Barcelona, Spain}
\affiliation[24]{Institute of Cosmology and Gravitation, University of Portsmouth, Dennis Sciama Building, Portsmouth, PO1 3FX, UK}
\affiliation[25]{Institute of Space Sciences, ICE-CSIC, Campus UAB, Carrer de Can Magrans s/n, 08913 Bellaterra, Barcelona, Spain}
\affiliation[26]{Fermi National Accelerator Laboratory, PO Box 500, Batavia, IL 60510, USA}

\affiliation[27]{School of Mathematics and Physics, University of Queensland, 4072, Australia}

\affiliation[28]{Departament de F\'{i}sica, Serra H\'{u}nter, Universitat Aut\`{o}noma de Barcelona, 08193 Bellaterra (Barcelona), Spain}

\affiliation[29]{Department of Physics and Astronomy, University of Sussex, Brighton BN1 9QH, U.K}

\affiliation[30]{Department of Physics \& Astronomy, University  of Wyoming, 1000 E. University, Dept.~3905, Laramie, WY 82071, USA}

\affiliation[31]{National Astronomical Observatories, Chinese Academy of Sciences, A20 Datun Rd., Chaoyang District, Beijing, 100012, P.R. China}

\affiliation[32]{Departamento de F\'{i}sica, Universidad de Guanajuato - DCI, C.P. 37150, Leon, Guanajuato, M\'{e}xico}

\affiliation[33]{Instituto Avanzado de Cosmolog\'{\i}a A.~C., San Marcos 11 - Atenas 202. Magdalena Contreras, 10720. Ciudad de M\'{e}xico, M\'{e}xico}

\affiliation[34]{Department of Physics and Astronomy, University of Waterloo, 200 University Ave W, Waterloo, ON N2L 3G1, Canada}

\affiliation[35]{Space Sciences Laboratory, University of California, Berkeley, 7 Gauss Way, Berkeley, CA  94720, USA}

\affiliation[36]{University of California, Berkeley, 110 Sproul Hall \#5800 Berkeley, CA 94720, USA}

\affiliation[37]{Instituto de Astrof\'{i}sica de Andaluc\'{i}a (CSIC), Glorieta de la Astronom\'{i}a, s/n, E-18008 Granada, Spain}

\affiliation[38]{Department of Physics and Astronomy, Sejong University, Seoul, 143-747, Korea}

\affiliation[39]{CIEMAT, Avenida Complutense 40, E-28040 Madrid, Spain}

\affiliation[40]{Department of Physics, University of Michigan, Ann Arbor, MI 48109, USA}
\affiliation[41]{University of Michigan, Ann Arbor, MI 48109, USA}

%% file: appendix.tex
\section{Computational Resources}
\label{computational-resources}

\subsection{Computational resources for `blobs'}
\label{sec:blob-resource}
Source measurements are segmented into units of `blobs'. The computational resources for each `blob' increase sharply with the area and number of sources in each `blob'. `Blobs' containing bright stars and galaxies often correspond to a large blob that contains many sources and is computationally expensive to process. Given the areas will not later be used for analysis, the \texttt{Obiwan} pipeline is made more efficient by ignoring them. The \textsc{Legacypipe} code produces a bitmask map for each brick; we exclude areas that are near optically bright and ``medium-bright'' stars, WISE $W1$-bright stars, large galaxies, and globular clusters.  These correspond to the ``maskbits'' data product bits 1, 11, 8, 12, and 13, respectively.

\subsection{Cosmos Deep}
\label{computational-resources-deep}
Because of the large number of exposures used in the COSMOS Deep production run, the computational time used to process these data in the same footprint is much higher than that in the DR9 production run. For example, brick 1503p015 uses 113 hours of wall time in the COSMOS Deep production run on one CPU node on NERSC's Cori machine, while in the DR9 production run it uses 8 hours of wall time on one node. The maximum numbers of overlapping exposures used in the DR9 production run in this region was 6, 7, 7 for $g,r,z$ bands, respectively. Meanwhile, the same numbers for the COSMOS deep production run are 45, 52, 120. This results in a much deeper depth in the three optical bands: The maximum \texttt{psfdepth} for the DR9 production run in this region are 25.19, 24.65, and 23.80 for
$g$, $r$ and $z$, while the median depth are  22.99, 23.94, 24.47 for the same bands. The corresponding numbers for the COSMOS Deep production run are 26.53, 26.27, 25.73 for maximum depth, and 25.31, 25.53, 24.79 for median depth. 

\subsection{Obiwan runs}
\label{appexdix:resources-obiwan}
\begin{figure}
    \centering
    \includegraphics[width=8cm]{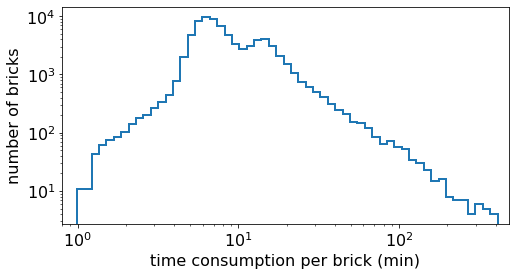}
    \caption{\texttt{Obiwan} simulation time consumption for bricks in the South Galactic Cap (SGC). One brick spans an area of 0.0625 square degrees. The median time to process one brick in the SGC is 7.56 minutes, and the average time consumed per brick is 10.89 minutes. }
    \label{fig:time-consumption}
\end{figure}

Figure \ref{fig:time-consumption} shows the time consumed for each brick. \texttt{Obiwan} runs on Permutter nodes at NERSC. With 128 threads on a single node, one node can process 8 bricks in parallel. On average it takes 11 minutes to finish one brick with 16 threads. Despite the fast speed of each brick, the pipeline is still inefficient. The problem is that the parallelization is limited by the memory (RAM) of the CPU available during the simulation. To read all the images needed for processing, we can only process a maximum of 8 bricks at a time. This means that with 16 threads available for computation, only 1 thread is active most of the time. The parallelization happens in the source fitting stage. As \texttt{Obiwan} fits many fewer sources than a normal reduction run, the efficiency on a per-source basis is therefore lower compared to the full image processing.

\subsection{Methods for Increasing the Efficiency of Synthetic Galaxy Production}
\label{appendix:speedup}
The current procedure of \texttt{Obiwan} is computationally expensive.  However, this is still feasible with image simulations significantly faster than \texttt{Obiwan}. One idea that allows significant speed up in estimating and mitigating galaxies at one location repeatedly. This would avoid repeating time-consuming stages like reading images, detecting sources, etc, and focus on the stage of source fitting. Furthermore, the procedure of source fitting can be accelerated by operating the code on GPUs, as it involves intensive utilization of matrix operations. 

Another practical way to address this problem is to develop a Monte Carlo procedure to estimate how galaxies are scattered under certain conditions analytically. This procedure numerically estimates how galaxy magnitudes are altered by estimating errors generated by each survey property. It is much faster than image simulation. Such a method would need to be calibrated on image simulation to validate the accuracy of its numerical estimations. The \texttt{Obiwan} results we have presented could be used for such calibration.

\section{Extra Details on Cosmos Deep}
\subsection{cut on $z$-band depth}
\label{appendix:z-band-cut}
\begin{figure}
\centering
\includegraphics[width=.48\linewidth]{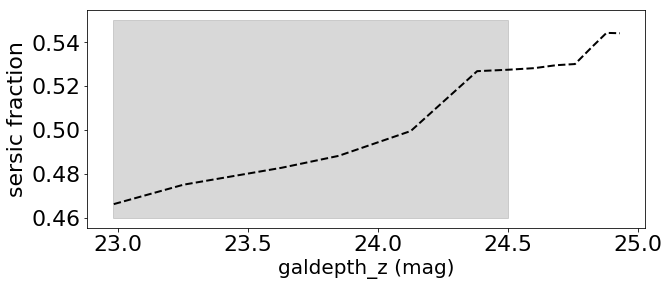}
\hfill 
\includegraphics[width=.48\linewidth]
  {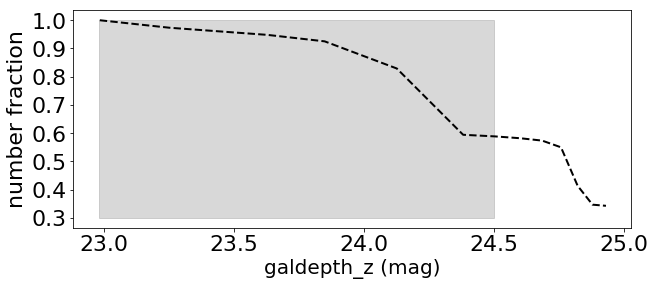}
\caption{Depth cut on the COSMOS Deep catalog. The left plot shows the fraction of \sersic\ galaxies when using different $z$-band galaxy depth (\texttt{galdepth\_z}) cuts.  The right plot shows the fraction of the remaining COSMOS deep catalog after the \texttt{galdepth\_z} cut. The increase of the \texttt{galdepth\_z} cutting limit will yield a higher fraction of \sersic\ galaxies in the LRG SV3 sample, and at the same time, there will be a decrease in the number of the total sample selected. We select a \texttt{galdepth\_z} cut of 24.5 as a trade-off between these two considerations.}
\label{fig:depthzcut}
\end{figure}

The depth of the COSMOS Deep Catalog allows us to use these data as a ``truth'' catalog, and our injected targets are sampled from it. However, even within the COSMOS Deep Catalog, there are significant variations in the depth and we first determine a minimum depth cut for the truth sample. The \texttt{Legacypipe} works in such a way that the complexity of the light profile chosen to obtain the photometry of a given source depends highly on the depth, based on simple criteria on the $\chi^2$ improvement required to allow a more complex profile (shown in equation \ref{equ:chi2-model}). We must make a trade-off in the truth catalog between its precision in shape, and its total galaxy counts. The ``truth input'' needs to be accurate enough in flux so that the output color distribution in simulated galaxies matches the real galaxies. Section \ref{sec:correlation-with-galaxy-brigntness} explains that the noise within the truth input leads to a mismatch in magnitude between real and synthetic LRGs. Figure \ref{fig:depthzcut} shows how these two variables compete given different depth cuts. We use the \texttt{galdepth} metric produced by \texttt{Legacypipe}, which gives the sensitivity to a round exponential-profile galaxy with a half-light radius of 0.45". We determine the depth cut by checking the percentage of \sersic\ galaxies in the sample. A higher fraction of \sersic\ galaxies in the sample means that the galaxies are better resolved in morphology. The determination of the type of a galaxy in \texttt{Legacypipe} has a prior that favors a simpler type (further explained in Section \ref{sec:truth-data-generation}). Because of this prior, the images have to be clear enough for the pipeline to choose the type \sersic\ for the galaxies in the images. By selecting regions with a higher fraction of \sersic\ galaxies, we have a higher confidence in the truth sample's LRG morphology, and the sample's photometry will also be better. Based on the observed distribution of \sersic\ fraction, we selected all data with \texttt{galdepth\_z} $>24.5$ into our truth sample.

\subsection{The WISE band in COSMOS Deep catalog}
\label{appendix:wise-deep}
\begin{figure}
\centering
\includegraphics[width=.35\linewidth]{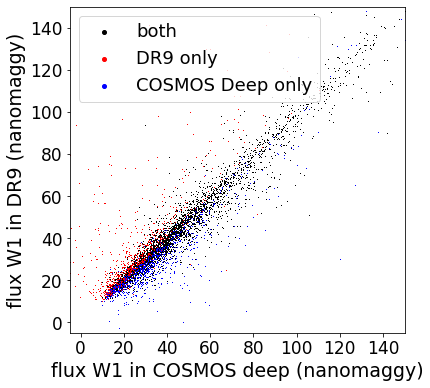}
\hfill 
\includegraphics[width=.64\linewidth]
  {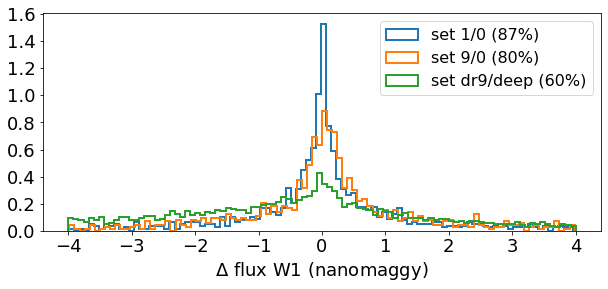}
\caption{(Left) Flux W1 for sources that pass LRG sv3 color cut on either DR9 catalogs or COSMOS deep catalogs. The black dots are sources passing LRG sv3 color cut on both sets, the red dots are sources passing LRG sv3 color cut only on the DR9 catalog. The blue dots are sources passing LRG sv3 color cut only on the COSMOS deep catalog. We see that most sources have similar W1 flux, and their differences in flux W1 value are likely due to photometric scattering. (right) Histogram of flux W1 difference distribution between two sets. Sources shown in this plot pass LRG sv3 color cut on either set. Blue is the difference between COSMOS repeat set 1 and 0. 87\% of the sources are within the range of in this plot. Orange is the difference between COSMOS set 9 and 0. 80\% of the sources are within the range of in this plot. Green is the difference between the DR9 catalog and the COSMOS deep catalog. 60\% are within the range of this plot. This plot suggests that flux W1 measured in COSMOS deep has a larger difference with DR9 catalog, compared with the difference that we see in between COSMOS repeat sets.}
\label{fig:depth_compare}
\end{figure}

We test the difference in the WISE flux measurements in the presence of different image sets in the optical band. We test it on 3 sets: the official DR9 release, the COSMOS Deep set, and the COSMOS Repeats set.  We compare the WISE flux on the same sources measured with different optical images. They are shown in Figure \ref{fig:depth_compare}. Figure \ref{fig:depth_compare} (left) suggests that there are not many false assignments of W1 flux to nearby sources since most sources in the two catalogs have similar W1 flux. However, Figure \ref{fig:depth_compare} (right) shows that the WISE flux dispersion between the COSMOS Deep and DR9 catalog is even bigger than between the COSMOS Repeats with the smallest and largest seeings (sets 0 and 9, respectively). One possible explanation is that the input catalogs from the optical bands have different number densities. The total number of sources detected mainly depends on \texttt{psfdepth}, as the extra faint sources detected are mostly point sources. For COSMOS repeat sets, despite having consistent \texttt{galdepth}, their \texttt{psfdepth} still increases with the set number. However, the difference between set 0 and set 9 is much smaller than the difference between DR9 and COSMOS deep. Thus, the difference we see for WISE flux in different sets is likely contributed by the difference in the number of sources given in the optical band. Potential errors in WISE PSF or morphological measurements in optical bands could induce uncertainty in flux W1 measurement, especially when the number of sources for "forced photometry" increases. However, no clear way exists to test which version has better precision for WISE flux measurement. We use the WISE flux from the COSMOS Deep measurement for consistency.

\section{De-biasing procedure for Cosmos Deep catalog}
\label{appendix:de-bias}
We compute the coadd \texttt{observed} image and the coadd \texttt{inverse\_variance} image, as well as the average \texttt{PSF} (point spread function) image of the COSMOS Deep data in each brick. Then we replace the coadd \texttt{observed} image with Gaussian noise based on pixel values from the \texttt{inverse\_variance} image. Next, we add one simulated COSMOS Deep source to its original pixel location, with the shape and flux information taken from the COSMOS Deep dataset. We define a blob that has this source centered in the middle of the image, with a size of $128 \times 128$ pixels. With these modifications, it is much faster for the pipeline to locate the designated blob, and fit the one galaxy that we injected. We then fit this blob using the same process in the \texttt{Legacypipe}, and \texttt{Legacypipe} returns the measured shape information. After we obtain the shape measured from \texttt{Legacypipe}, we compare it with the input shape. When they don't match, we modify the input shape and run the pipeline for more time. We repeat this process until the output shape reaches the desired precision. Although there is only a prior on ellipticity in equation \ref{equ:chi2-model}, we find that the parameters of galaxy radius and \sersic\ index are also biased when we check the shape distribution of \texttt{Obiwan} galaxies that directly samples from the COSMOS Deep catalog. It is likely because parameters for galaxy shapes are correlated with each other. We define a quantity \textbf{A} that includes 3 shape parameters: $e$ (ellipticity), $shape\_r$ (half-light radius), and $sersic\_n$ (\sersic\ index). It is defined by 

%TODO
\begin{equation}
\begin{split}
\textbf{A}_i &= (\texttt{sersic\_n}_i^O - \texttt{sersic\_n}_\texttt{s})^2/5.5 + (\texttt{e}_i^O - \texttt{e}_\texttt{s})^2 \\
&\quad + (\texttt{shape\_r}_i^O - \texttt{shape\_r}_\texttt{s})^2
\end{split}
\end{equation} 

The upper index O here denotes the output value returned by \texttt{Legacypipe}. The lower index i denotes the number of iterations. The lower index s denotes values taken from the COSMOS Deep dataset, and it is also the starting value of the iteration. Our goal is to have a similar shape distribution between the output \texttt{Obiwan LRG} catalog and the \texttt{DESI LRG} catalog. To achieve this, we require that the shape parameter returned by \texttt{Legacypipe} is similar to the ones measured by the COSMOS Deep set. When the condition is achieved, we record the input shape parameters to use them as input for injected galaxies. 

When the \textbf{A} value does not reach our desired accuracy, we modify the parameters. Each time the parameters are modified by:
\begin{equation}
\text{$X_{i+1}^I$} = \text{$X_i^I$} - (\text{$X^O_i$} - X_s)*step 
\end{equation}

The $X$ here denotes one of the parameters from $sersic\_n$, $e$ or $shape\_r$. The upper index "I" denotes the input value from i\emph{th} iteration. The \textit{step} is a constant number used to determine how much the new parameter is modified. The three parameters are all updated during each iteration. We added some constraints on the ellipticity $e$. We force galaxies with $e=0$ (round galaxies or point source) in the deep photometry to remain $e=0$. Galaxies with non-zero ellipticity preserve their orientation. 

These new set of parameters are accepted if 
\begin{equation}
\label{equ:chi}
\text{$\textbf{A}_{i+1}<\textbf{A}_i$}
\end{equation}

Otherwise, we make the $step$ size smaller and try again, with a maximum of 3 times per trial. The $step$ value is defined as 0.2, 0.1, 0.05 for 0 fail, 1 fail, and 2 fails. The $step$ value is chosen so that the code can reach the correct value quickly and precisely.
%In general, if we think about the relationship between $X_i^I$ and $X_i^O$, $X_i^O$ monotonically increases with $X_i^I$. Due to the prior in equation \ref{equ:chi2-model}, the slope for this monotonical function is less than 1. Thus, the $X^I$ we eventually get should conceptually be $(X_s - X_i^O )/(X^I - X_i^I)<1$, which translates to $X^I>X_i^I - (X_i^O-X_s)*1$. This means that $step<1$ is generally safe for the parameter to reach the right value. Thus, we let $step$ be 0.2, 0.1, 0.05 for three tails, which are all smaller than 1, and can give us enough precision to fine-tune on these parameters. 

There are two criteria to stop the iteration. First, when the equation fails 3 times, it means that the current value is the smallest \textbf{A} we can get. Second, If \textbf{A}< $10^{-4}$, it means that the error is small enough. On either occasion, we stop the iteration.

There are many advantages to a procedure like this. First, it makes debiasing computationally possible by working on the coadd images instead of individual exposures. A COSMOS Deep brick takes more than 40 hours to finish on NERSC's one Cori node. Since we need to fit on the same spot for each source involved iteratively, it would take even more time than a COSMOS Deep run. Second, it automatically takes into account the bias from blending, without actually being contaminated by nearby sources: The blended region is more noisy and has a lower value on \texttt{inverse\_variance} image. Also, when blending exists in a pixel, it would be brighter and has a higher Poisson noise.

\begin{figure}
\centering
\includegraphics[width=13cm]{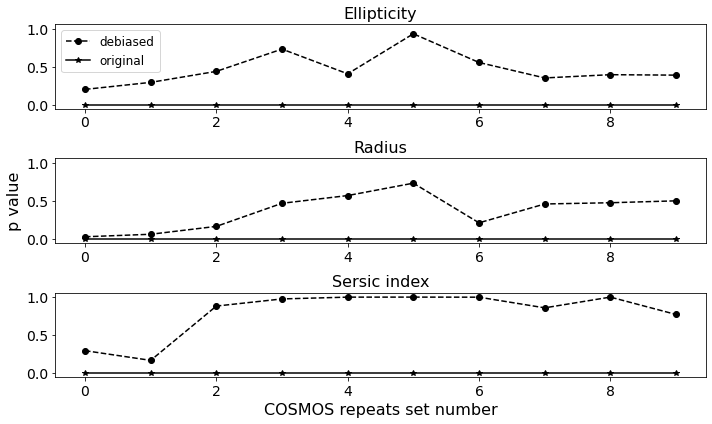}
\caption{Histogram comparison between the simulated LRGs and the real LRGs in the same COSMOS Repeats footprint. This plot's dot and star curves represent different types of simulated galaxies. The dot-dashed curve uses input seeds from the de-biased sample, and the star solid curve uses input seeds from the un-modified sample. The x-axis denotes the ten sets of the COSMOS Repeats numbered 0 to 9, and the y-axis uses a p-value to indicate how close the histograms are between the simulated and the real galaxy shape distribution, using the KS test, 1 being the most similar and 0 being the least similar. These plots compare the shape parameters of ellipticity, \sersic\ index, and the half-light radius.}
\label{fig:debias}
\end{figure}

We test our shape-modified set on the COSMOS Repeats set. We inject these shape-modified galaxies into COSMOS Repeats images and recover the output shape with \texttt{Obiwan}. We compare the shape distribution of our simulated LRGs with the real LRGs in the same set and perform the 
Kolmogorov–Smirnov (KS) test over the distribution; the results are shown in Figure \ref{fig:debias}. One can observe that based on the $p$-values of the KS test, the distribution of the shape-modified catalog (dot-dashed) is much more similar to the \texttt{DESI LRGs} when compared with the un-modified version (star-solid). The $p$-values of the un-modified data are near zero, thus clearly rejecting the possibility that the un-modified data are representative of \texttt{DESI LRGs}. Conversely, the $p$-values of the modified data increase in all cases and are acceptable in most, and thus
the debiasing process yields considerable improvement in the output shape distribution of \texttt{Obiwan LRGs}. 

\section{Measurement bias}
\label{appendix:measurement-bias}
\begin{figure}
\centering
\includegraphics[width=15cm]{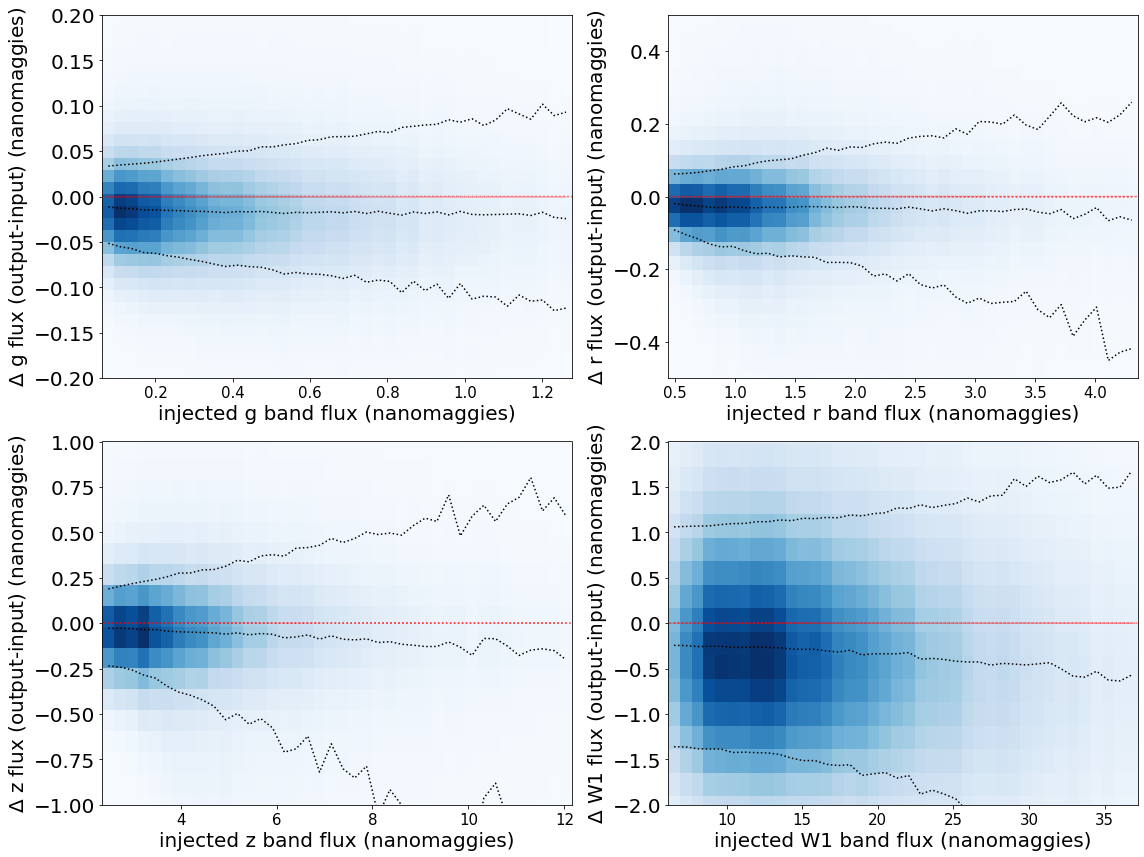}
\caption{The density (blue) plot of output flux minus input flux. The black dashed line is the 84th, 50th, and 16th percentile for a given input magnitude. The red dashed line is centered at 0 on the y-axis. The difference between the red line and the 50th-percentile line shows that the flux measurement is consistently biased at all input flux levels in all bands. The four plots show results in the $g,r,z, W1$ band, respectively.}
\label{fig:FluxBiasDensity}
\end{figure}

Overall, the flux measurement is negatively biased. Figure \ref{fig:FluxBiasDensity} shows the measurement flux bias as a function of input magnitude in each band. The dashed line in the middle is the median value in each flux bin. It shows that the bias is consistent across all magnitude bins, and it exists in all optical and infrared bands. To understand whether these features are a result of pipeline structures, or some features in the images, we designed four parallel runs: On the first run, we first generate synthetic Gaussian background noise that is much smaller than a typical Legacy Survey image. We inject galaxies into these images. In this setting, there are very few galaxies in the image, and no blending would occur here. The image is also almost noiseless. The second run is similar to the first run. The difference is that we set the Gaussian noise to the level of a typical Legacy Survey Image. In the third setting, we take the images on the second run and add the real galaxies back with their model parameters recorded in the DR9 data release. With this setting, we have a fully simulated image without any contamination. Blending may occur in such images, but it is free of error from background subtraction, cosmic ray contamination, etc. In the final setting, we use the real images and add simulated galaxies to these images, which is the same way as our previous production runs. The 4 production runs use the same input galaxy setting on the same set of images. In total, 442 bricks are processed for this comparison. Table \ref{tab:run-compare} shows the results of these production runs. The results show that flux bias does not appear on these non-real images. This means that there is no algorithmic negative bias in the pipeline. The bias from real images is caused by other effects that can not be easily simulated. Fortunately, the bias is small enough that it has little effect on how we select galaxies as LRGs. If we correct the bias in our simulated outputs as a function of flux: 
\begin{align}
    {}& \mathit{flux\_g}_{\mathit{corrected}} = 1.0123*\mathit{flux\_g}_{\mathit{measured}}  + 0.0087 \\
    {}& \mathit{flux\_r}_{\mathit{corrected}} = 1.0112*\mathit{flux\_r}_{\mathit{measured}}  + 0.0100  \\
    {}& \mathit{flux\_z}_{\mathit{corrected}} = 1.0166*\mathit{flux\_z}_{\mathit{measured}}  + 0.0276 \\
    {}& \mathit{flux\_{W1}}_{\mathit{corrected}} = 1.0070*\mathit{flux\_{W1}}_{\mathit{measured}} + 0.2483
\end{align}

The population of galaxies that pass the target selection cuts with the new flux values only changes by 0.77\%. Therefore, this bias has a negligible effect on determining the target selection of LRGs.

\begin{table}
\begin{center}
\begin{tabular}{||c| c| c| c|c|} 
 \hline
 \diagbox[width=10em]{median value}{images} & negligible noise & regular noise & Noise + model galaxies &  Real images\\ [0.5ex] 
 \hline\hline
 $\Delta \mathit{flux\_g}$ (nanomaggies) & 0.005 & 0.013 & 0.017 & -0.014 \\ 
 \hline
  $\Delta \mathit{flux\_r}$ (nanomaggies) & 0.002 & 0.017 & 0.024 & -0.03 \\
 \hline
  $\Delta \mathit{flux\_z}$ (nanomaggies) & -0.009 & 0.057 & 0.062 & -0.063 \\
 \hline
\end{tabular}
\end{center}
\caption{Median flux difference (output-input) for 4 production runs. In the first column,
simulated galaxies are added to images containing tiny Gaussian noise (standard deviation of $10^{-4}$ nanomaggies per pixel). In the second column, simulated galaxies are added to images with Gaussian noise expected in the weight image. In the third column, galaxies are added to images with regular-level Gaussian noise, plus idealized model galaxy images for the DR9 catalog.  In the fourth column, simulated galaxies are added to real DR9 images, as in our main \texttt{Obiwan} runs.  In the real images, the fluxes are consistently biased low. This does not happen in other image sets. For the "negligible noise" set, there is approximately no bias present. In the "regular noise" and "Noise + model galaxies" sets, the flux values are biased high. }
\label{tab:run-compare}
\end{table}

\section{Testing impact of \texttt{psfsize} with COSMOS repeats}
\label{appendix:psfsize}
\subsection{Origination of trend with \texttt{psfsize}}
 \label{sec:psf_study}
 The COSMOS Repeats data (described in Section \ref{sec:cosmos-repeat}) is suitable for this analysis because of its shallow and consistent \texttt{galdepth} and varying seeing (\texttt{psfsize}). We use COSMOS  Deep as truth data for COSMOS Repeats as COSMOS Deep fully covers the region of COSMOS Repeats. 
 
\begin{figure}
    \centering
    \includegraphics[width=0.48\columnwidth]{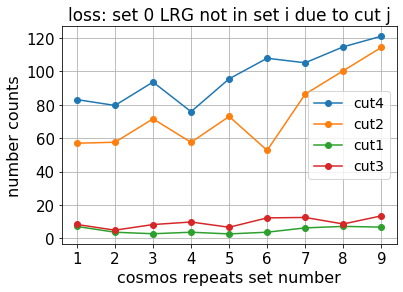}
    \hfill
    \includegraphics[width=0.48\columnwidth]{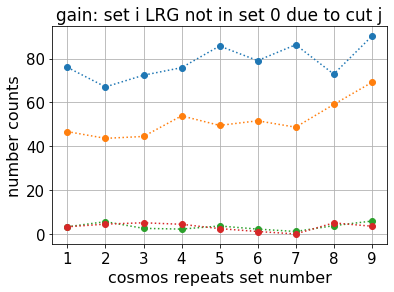}
    \caption{Number counts of LRG/LRG-like sources scattering outside or inside each color cut boundary. The left plot shows the number counts of LRGs in COSMOS Repeat set 0, but not in COSMOS Repeat set $i$ ($i$ is from 1 to 9), and does not pass the color cut $j$ ($j$ is from 1 to 4). We call these LRGs ``lost'' with respect to set $i$. The right plot shows the number counts of LRGs in COSMOS Repeat set $i$, but not in COSMOS Repeat set 0, and does not pass the color cut $j$. We call these LRGs ``gained'' with respect to set $i$. The cuts 1--4 are the same ordering as in equations \ref{equ:lrg-like-1}--\ref{equ:lrg-like-4}. Two of the four cuts dominate the photometric scattering. Cut 2 is the ``faint limit'' cut on $z$ band fiber magnitude. Cut 4 is the ``luminosity cut'' to select LRGs at higher redshift \cite{zhou2023target}. LRGs around these two cuts tend to have low magnitudes, so there is more photometric scattering around these two cuts.}
    \label{fig:RepeatGainLoss}. 
\end{figure}

  We first test which LRG color cuts have large scattering among the 4 equations \ref{equ:lrg-like-1}--\ref{equ:lrg-like-4}. In Figure \ref{fig:RepeatGainLoss}, We use COSMOS Repeats set 0 as a reference set. For any set $i$ in sets 1 through 9, we compute the number counts of ``Loss'' and ``Gain''. ``Loss'' is defined as sources passing LRG color cut in reference set 0, but not in set $i$. We consider them ``lost LRGs'' in set $i$. For each color cut, we compute the number counts of ``lost LRGs'' that do not pass this cut. ``Gain'' is the opposite of ``Loss'': They are the LRGs that pass the LRG color cut in set $i$, but not in set 0. We consider set $i$ ``gained'' these LRGs. We check the number counts of these ``gained LRGs'' that do not pass a certain color cut in set 0. Cuts with high number counts in both the ``loss'' and ``gain'' plots are the dominant power to produce photometric scattering and, thus, potential density fluctuations. Here, we see that cut 2 (equation \ref{equ:fibermagz}) and cut 4 (equation \ref{equ:lrg-like-4}) have this dominant power in producing density fluctuation. Thus, we focus on these two cuts in the following analysis. 

\begin{figure}
    \centering
    \includegraphics[width=0.48\columnwidth]{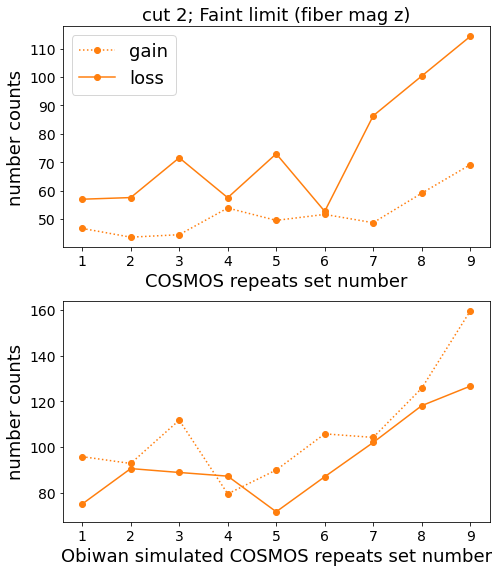}
    \hfill
    \includegraphics[width=0.48\columnwidth]{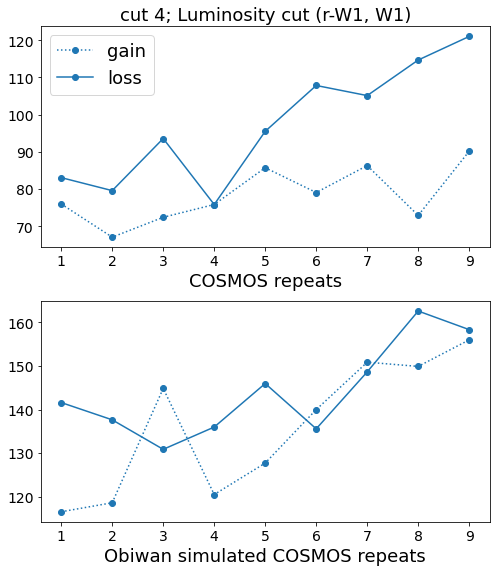}
    \caption{Comparison for LRGs ``gained'' and ``lost'' (described in Figure \ref{fig:RepeatGainLoss}) for cut 2 (left, equation \ref{equ:fibermagz}) and cut 4 (right, equation \ref{equ:lrg-like-4}). The upper plot shows results from the original COSMOS Repeats sets. The lower plots show the result from Obiwan simulations with the same setting. The ``gain'' and ``loss'' curves in Obiwan simulations do not deviate from each other. Meanwhile, LRGs in the original COSMOS Repeats sets show deviation between ``gain'' and ``loss'' starting from set 6 and beyond. The result suggests signs of LRG density variation from seeing (\texttt{psfsize}) that is not simulated by \texttt{Obiwan}. We see a net loss of LRGs at high seeing that is not simulated by \texttt{Obiwan}, which is consistent with the trend seen in Figure \ref{fig:psfsize_z_mismatch}, where we see a downward trend of LRGs as a function of \texttt{psfsize}. }
    %and figure \ref{fig:psfsize_splits}.
    \label{fig:cosmos_deviation}
\end{figure}

Density variation appears when number counts in ``loss'' and ``gain'' deviate from each other. In Figure \ref{fig:cosmos_deviation}, we see that the COSMOS Repeats show such deviation starting from COSMOS Repeat set 6. However, \texttt{Obiwan LRGs} simulated in the same setting do not show a clear deviation in all sets. This is a sign that some unknown factor not simulated by \texttt{Obiwan} is driving the deviation we see in the real COSMOS Repeats sets. As cut 2 (equation \ref{equ:fibermagz}) is a simple cut on fiber magnitude $z$, it is easier to investigate. It is also possible to investigate cut 4 (equation \ref{equ:lrg-like-4}), but due to its complicated functional form, we focus on testing cut 2 in the following paragraphs. 

\begin{figure}
    \centering
    \includegraphics[width=0.6\columnwidth]{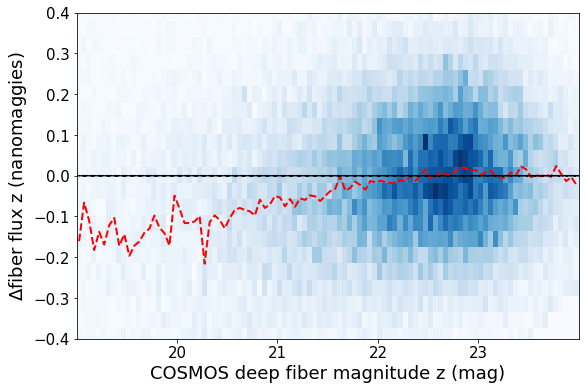}
    \caption{Flux difference between \textbf{all} sources measured in the COSMOS Repeats set 9 catalog, and sources measured in the COSMOS Deep catalog, as a function of fiber $z$-band magnitude in the COSMOS Deep catalog. The blue patches are the density of sources. The red line is the median value of $\Delta$fiber flux as a function of $z$-band fiber magnitude in COSMOS Deep. Sources in the COSMOS Repeats set 9 catalog are statistically fainter than their counterparts in the COSMOS Deep catalog below \texttt{fiber z magnitude} 22.}
    \label{fig:fiber_z_bias}
\end{figure}

\begin{figure}
    \centering
    \includegraphics[width=0.6\columnwidth]{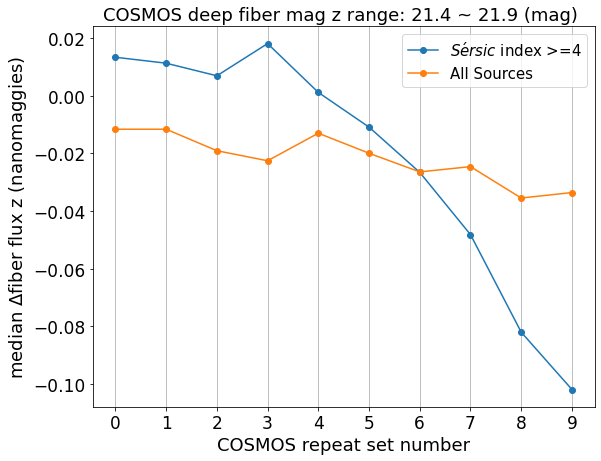}
    \caption{Median $\Delta$fiber flux between the COSMOS Repeats set $i$ ($i$ from 1 to 9) catalogs and the COSMOS Deep catalog, with the COSMOS Deep catalog having a fiber $z$ magnitude range of 21.4 to 21.9. Orange is for all sources, while blue is for sources with \sersic\, index >=4. We chose this range because this region is sensitive to LRG color cut \ref{equ:fibermagz}. With increased seeing, flux measured in the COSMOS Repeats set is statistically fainter. The trend is prominent for more extended sources. This could be an explanation for fewer LRGs observed in regions with worse seeing.} 
    \label{fig:cosmos_repeats_psfsize_trend}
\end{figure}

We first looked at the histogram of fiber mag $z$ distribution of LRGs in a counterpart set that failed this color cut. The distribution looks like normal photometric scattering: Their values are close to the fiber mag $z$ cut of $21.7$. Thus, we suspect that the seeing (\texttt{psfsize}) is making the sample consistently fainter or brighter. Figure \ref{fig:fiber_z_bias} validates our assumption. We match \textbf{all} sources in COSMOS Repeats set 9 to the COSMOS Deep data. We find that at \texttt{fiber z magnitude < 22} in COSMOS Deep, the flux in COSMOS Repeats set 9 is statistically lower. For sources around \texttt{fiber z magnitude = 21.7}, the scattering range is about \texttt{fiber z magnitude} around $21.4$ to $21.9$. We checked the median \texttt{fiber z flux} difference between each COSMOS Repeats set and COSMOS Deep. Figure \ref{fig:cosmos_repeats_psfsize_trend} shows the median $\Delta$flux value in all sets. There is a clear downward trend from set 0 to set 9, meaning that the higher seeing is causing the measured \texttt{fiber z flux} around $21.4$ to $21.9$ to appear fainter. Notably, this trend is greatly amplified when we exclusively choose more extended sources. When we look at sources with \sersic\, index >= 4, we find this trend to be much larger. According to COSMOS deep data, 43\% of \texttt{DESI LRGs} have \sersic\, index >=4, so this does have a strong impact. This process effectively shifts cut 2 (equation \ref{equ:fibermagz}) to a higher threshold for cutting the true fiber flux value, as the observed flux is statistically lower than the true flux. The more ``loss'' of LRGs in high-seeing regions is a result of this shift. As it excludes a small fraction of LRGs close to the \texttt{fiber z magnitude = 21.7} color cut.

A potential shift could also appear round cut 4 (equation \ref{equ:lrg-like-4}). We can not test it due to the complexity of the color-cut equation. COSMOS repeat set is a small dataset and it is difficult to use it to test a color cut that involves two bands ($r$-band and $W1$-band). However, we do see that LRGs close to the \texttt{fiber z magnitude = 21.7} boundary are also close to the boundary in cut 4 (equation \ref{equ:lrg-like-4}). Thus, potential density variations resulting from shifts around cut 4 (equation \ref{equ:lrg-like-4}) could also introduce variation of LRGs around \texttt{mag z = 21} at different seeing as shown in Figure \ref{fig:psfsize_splits}.

We conclude that we find signs of PSF errors not simulated by \texttt{Obiwan}. This error shifts the statistical distribution of sources to a slightly fainter flux in regions of worse seeing. This shift induces a loss in the number density of LRGs. It could also be the underlying reason for the density dispersion of LRGs around \texttt{mag z = 21} at different seeing that we see in Figure \ref{fig:psfsize_splits}.

\subsection{Morphological change of LRGs with \texttt{psfsize}}
Two phenomena are going on in Figure \ref{fig:psfsize_splits}. The first phenomenon is that with an increase of \texttt{psfsize z}, there is an increase in the LRG density at the faintest magnitudes, $z> 21.3$. This phenomenon is replicated in \texttt{Obiwan LRGs}. This phenomenon is likely because there is more photometric scattering at high \texttt{psfsize z} as these regions typically have lower depth. Sources in this \texttt{mag z} > 21.3 regions with the \texttt{fiber mag z} < 21.7 cuts are compact sources. According to results from \texttt{Obiwan}, their true color is more likely to be outside LRG color selection boundaries. The second phenomenon is described in Section \ref{sec:psf_study}, which causes variation at around \texttt{mag z = 21}. The sources lost due to this phenomenon likely have a more extended shape because they have a more significant difference between \texttt{mag z} and \texttt{fiber mag z}. 

\begin{figure}
    \centering
    \includegraphics[width=0.48\columnwidth]{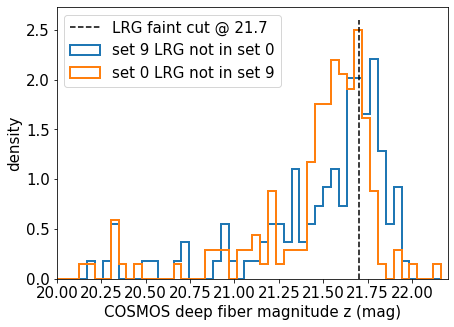}
    \hfill
    \includegraphics[width=0.48\columnwidth]{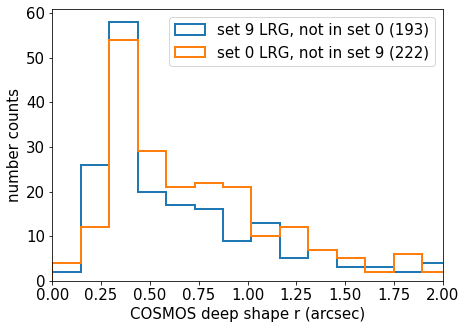}
    \caption{Histogram of fiber magnitude $z$ (left) and half-light radius of the galaxy model for galaxies (right) in the COSMOS Deep catalog for LRGs in COSMOS Repeats set 9 but not in set 0 (blue), and LRGs in COSMOS Repeats set 0 but not in set 9 (orange). The numbers in the legend of the right plot denote the total count of sources plotted in each histogram. The black dashed line in the left plot is the LRG color cut of \texttt{fiber magnitude} $< 21.7$. 31\% of LRGs in set 9 but not in set 0 are outside the LRG faint cut, while 20\% of LRGs in set 0 but not in set 9 are outside the LRG faint cut. They suggest that LRGs gained in a better seeing are statistically brighter. The right plot shows that LRGs gained in a better seeing have a larger half-light radius, which suggests that LRGs gained in a better seeing have more extended profiles.}
    \label{fig:COSMOS_LRG_in_deep}
\end{figure}

The sources gained in the first phenomenon and the sources lost in the second phenomenon are not likely to have a large overlap because of their difference in morphology. Figure \ref{fig:COSMOS_LRG_in_deep} shows the COSMOS Deep fiber magnitude and half-light radius of LRGs selected \textbf{only} in COSMOS Repeats set 0 or set 9. Set 0 LRGs are statistically brighter and have a more extended profile than set 9 LRGs. This difference validates our claim that LRGs have a slightly different population at different seeing. For \texttt{DESI LRGs}, we may not see a large density variation in seeing as the ``gained'' and ``lost'' in these two phenomena add up. However, it could pose potential risks in cosmological analysis as the two populations of LRGs have different sample statistics. Compact galaxies have less star-forming processing going on than extended galaxies. Even if they are similar in photometric color, they may differ in stellar mass, which results in a difference in galaxy bias.

